\documentclass[iop,numberedappendix]{emulateapj}
\usepackage{natbib}
\usepackage{amsmath}

\newcommand{\msol}{\,\textrm{M}_\sun}  

\shorttitle{The Dark Matter Dist. in Low-Mass Disk Galaxies: II}
\shortauthors{Relatores et al.}

\begin{document}
\title{The Dark Matter Distributions in Low-Mass Disk Galaxies. II. The Inner Density Profiles}
\author{Nicole C. Relatores$^{1,2}$, Andrew B. Newman$^{2,1}$, Joshua D. Simon$^2$, Richard S. Ellis$^3$, Phuongmai Truong$^4$, Leo Blitz$^4$, Alberto Bolatto$^5$, Christopher Martin$^6$, Matt Matuszewski$^6$, Patrick Morrissey$^6$, James D. Neill$^6$}
\affil{$^1$ Department of Physics and Astronomy, University of Southern California, Los Angeles, CA, 90089-0484, USA\\
$^2$ Observatories of the Carnegie Institution for Science, Pasadena, CA 91101, USA\\
$^3$ Department of Physics and Astronomy, University College London, Gower Street, London WC1E 6BT, UK\\
$^4$ Department of Astronomy, University of California, Berkeley, CA 94720, USA\\
$^5$ Department of Astronomy, University of Maryland, College Park, MD 20742-2421, USA\\
$^6$Cahill Center for Astrophysics, California Institute of Technology, 1216 East California Boulevard, Mail code 278-17, Pasadena, CA 91125, USA}

\begin{abstract}
Dark matter-only simulations predict that dark matter halos have steep, cuspy inner density profiles, while observations of dwarf galaxies find a range of inner slopes that are often much shallower. There is debate whether this discrepancy can be explained by baryonic feedback or if it may require modified dark matter models. In Paper 1 of this series, we obtained high-resolution integral field H$\alpha$ observations for 26 dwarf galaxies with $M_*=10^{8.1}-10^{9.7}\msol$. We derived rotation curves from our observations, which we use here to construct mass models. We model the total mass distribution as the sum of a generalized Navarro-Frenk-White (NFW) dark matter halo and the stellar and gaseous components. Our analysis of the slope of the dark matter density profile focuses on the inner 300-800 pc, chosen based on the resolution of our data and the region resolved by modern hydrodynamical simulations. The inner slope measured using ionized and molecular gas tracers is consistent, and it is additionally robust to the choice of stellar mass-to-light ratio. We find a range of dark matter profiles, including both cored and cuspy slopes, with an average of $\rho_{\rm DM}\sim r^{-0.74\pm 0.07}$, shallower than the NFW profile, but steeper than those typically observed for lower-mass galaxies with $M_*\sim 10^{7.5}\msol$. Simulations that reproduce the observed slopes in those lower-mass galaxies also produce slopes that are too shallow for galaxies in our mass range. We therefore conclude that supernova feedback models do not yet provide a fully satisfactory explanation for the observed trend in dark matter slopes.  

\end{abstract}

\section{Introduction}\label{Intro}
Explaining the distribution of dark matter on large scales is a key success of the $\Lambda$ cold dark matter ($\Lambda$CDM) model, but several important questions remain when considering smaller scales (\citealt{Bullock}). In particular, the distribution of dark matter in low-mass galaxies has long been viewed as a challenge to the $\Lambda$CDM model, and it remains to be determined if this discrepancy can be resolved by accounting for baryonic effects or if it is a result of dark matter microphysics that we have yet to understand. 

\cite{NFW} used dark matter-only N-body simulations to show that a single universal density profile can be used to describe dark matter halos across a wide range of mass scales. This Navarro-Frenk-White (NFW) profile has been confirmed by observations on large scales (e.g. \citealt{Rachel,Umetsu}), but some differences emerge on sub-galactic scales, in particular in the central regions of low-mass galaxies. 


The NFW profile dictates that at small radii, the dark matter density scales as $\rho_{\rm DM} \propto r^{-1}$, or what is known as a `cusp'. Instead, observations of dwarf galaxies have found these galaxies often host shallower density profiles, generally called `cores', where at small $r$ the dark matter distribution scales as $\rho_{\rm DM} \propto r^{-\beta}$, with $\beta \sim 0$ \citep{deblok01-2, deblok01, blokbosma, Josh03, kuzio08, Oh2011t, Oh2015}. This `cusp-core' problem was first noticed in low-mass dwarf galaxies (e.g., \citealt{FandP, Moore}), which are dark matter dominated and so are excellent laboratories for studying dark matter. 

The apparent discrepancy between theory and observations can possibly be attributed to the fact that the simulations used to find the NFW profile were dark matter-only, and therefore did not incorporate any kind of baryonic physics. Modern hydrodynamical simulations now include baryons in their models, and results from these simulations indicate that baryonic feedback may be responsible for flattening the inner dark matter density profile. In particular, feedback from supernovae could redistribute dark matter in the central regions through the change induced in the gravitational potential by expelled gas (e.g., \citealt{NFW-feedback, Governato}). Repeated fluctuations of the central potential over time will move the dark matter irreversibly outward, leading to a reduced central density. In order to be effective, this process requires the frequent, repeated presence of gas outflows induced by bursts of star formation \citep{Sergey,Pontzen,C-Cfb}. 

On the other hand, modifying our model of cold dark matter could also account for some of the differences between theory and observation. Warm dark matter does not seem to be a likely candidate as collisionless dark matter that is warm enough to create cores is in conflict with other observations \citep{KuzioWDM, Maccio, Shao}. Scalar field dark matter models prove complicated to simulate and cannot yet account for the flattening of the inner dark matter profile \citep{Zhang, Bernal}. Allowing self-interactions between dark matter particles, however, can create cores in dwarf galaxies and has not been ruled out by observations \citep{Spergel, Kaplinghat, Harvey,  Robertson}. Self-interacting dark matter (SIDM) would have significant effects only where the dark matter density is sufficiently high, thus preserving the large-scale successes of the $\Lambda$CDM model. Additionally, SIDM would create a coupling between the dark matter and the baryons due to the thermalization of the inner halo, naturally leading to a diversity of dark matter density profiles (\citealt{Kamada}). 

While plausible solutions to the cusp-core problem exist, in order to determine the process(es) responsible for creating shallow inner dark matter density slopes, we require a better understanding of the distribution of the slopes found in nature. This will help us to determine if every dwarf galaxy deviates from NFW, if the value of the inner dark matter slope is related to other galaxy properties, and if feedback or SIDM models can accurately describe the observational results. Achieving this goal requires a large sample of low-mass galaxies with high quality observational data. 

For this reason, we have obtained high-resolution H$\alpha$ kinematics for a sample of 26 dwarf galaxies using the Palomar Cosmic Web Imager (PCWI; \citealt{CWI}). Paper 1 of this series \citep{Paper1} details our galaxy sample ($\log L_r/\textrm L_{\sun}= 8.4-9.8$, $v_{\rm max} = 50-140$ km s$^{-1}$) and observations with PCWI. Although larger sets of rotation curves have been compiled from \ion{H}{1} and H$\alpha$ observations (e.g., the SPARC compilation; \citealt{sparc}), ours is among the largest with two-dimensional kinematic data, which allowed us to assess the level of non-circular motions (Paper 1, Section 4). We also sample the inner kpc better than most literature rotation curves (3-6 points versus typically 0-2 in the SPARC dwarfs), which is important to discriminate physical models of the dark matter distribution \citep{dicintio16}.

This survey was conducted alongside a similar investigation using CO observations from the Combined Array for Research in Millimeter-wave Astronomy (CARMA; \citealt{Mai1}), and for a subset of 11 galaxies we compared velocity fields derived from the different tracers, finding only small differences on the scale of the random motions of the interstellar medium (ISM). Paper 1 also details the derivation of rotation curves from the H$\alpha$ velocity fields.  

In this paper, we use those rotation curves to construct models of the mass distribution of each galaxy. We model the rotation velocities as the sum of the contributions from the stellar, gaseous, and dark matter components. Using previously obtained optical and infrared photometry, we measure the stellar component, while for a subset of galaxies, we use the CO observations from CARMA to probe the molecular gas distribution, which typically dominates the gas mass in the inner few kiloparsecs. This allows us to construct a model of the dark matter in our galaxies, and thus infer the slope of the density profile in the central region, from which we can assess any deviations from NFW that are present. Due to the size of our sample, we are able to examine the distribution of dark matter slopes as well as search for correlations with the baryon distributions. 

This paper is organized as follows: Section 2 details our measurements of the stellar distribution from photometry, Section 3 describes our measurements of the gas mass distributions, Section 4 describes the mass modeling process, Section 5 provides robustness tests for our models, Section 6 gives our results on the inner dark matter density profiles, Section 7 discusses our work in the context of other observational work and simulations, and finally Section 8 summarizes our results.


\section{Measuring the Stellar Mass Distribution}\label{stars}

\subsection{Stellar Contribution to the Rotation Curves}\label{MGE}
The rotation curves derived in Paper 1 are a result of the gravitational pull of both the baryonic matter and dark matter in each galaxy.  To isolate the effect of the baryons from that of the dark matter, we need to determine how much the stars and gas contribute to the rotation curve. We will discuss the effects due to the stars only in this section, while the discussion of gas follows in Section \ref{gas}. 

To estimate the potential generated by the stellar distribution we use the package presented by \cite{MGE}, which gives an accurate and robust algorithm for determining multi-Gaussian expansion (MGE) fits to galaxy images. We use images taken in the $r$-band for all galaxies in our sample, in addition to infrared (4.5 $\mu$m) images for the 18 galaxies that have archival data from Channel 2 of Spitzer's IRAC (details about the imaging data can be found in Paper 1). When both are available, we preferentially use the infrared images, as they more closely trace the stellar mass distribution. The center, PA, and ellipticity of each galaxy are kept fixed to the values determined in Section 3.2 of Paper 1 (see below for exceptions). Note that our sample was restricted to galaxies with inclinations between $30^{\circ} -70^{\circ}$, as galaxies close to edge on would not allow us to see the internal motions, and those close to face on would prevent us from measuring the rotation. The MGE package fits a series of Gaussians to the image in order to produce a surface brightness profile and a luminosity. The stellar contributions to the rotation curves are then found using Jeans Anisotropic Modeling, as described in \cite{JAM}, to calculate the circular velocity in the potential generated by the stars. We model the stellar disk as an oblate ellipsoid with thickness $c/a = 0.14$ \citep{qmin}.  

As mentioned, during the MGE fitting process, the galaxy geometry is kept fixed to values determined in Paper 1. For NGC 959 and NGC 7320, we observed a monotonic decrease in ellipticity towards the center, potentially indicating the presence of a bulge. Since this could modestly influence the rotation curve, in these two cases we allowed the ellipticity to vary with radius for the MGE fits.

\subsection{Photometric Estimates of the Stellar Mass-to-Light Ratio}\label{ML}
The above procedure traces the luminous stellar distributions and therefore the shape of the stellar contribution to the rotation curve. However stellar population synthesis (SPS) models are needed to estimate the stellar mass, and therefore the amplitude of the stellar contribution to the rotation curve. In this section, we use SPS models to give us a plausible estimate of the stellar mass-to-light ratio $\Upsilon_*$ for each galaxy. Due to the uncertainty in these models and the possible dependence of the outcome on model choice, we will explore the effect of different stellar mass-to-light ratios on the inferred dark matter density profile in Section \ref{ML_robust}. 

The NASA-Sloan Atlas (NSA; \citealt{NSA}) contains 17 of the galaxies in our sample. The information in this catalog allows us to investigate the mass-to-light ratio with homogeneous, matched-aperture photometry in the $ugriz$ bands. 

Fitting and Assessment of Synthetic Templates (FAST), introduced by \cite{FAST}, is a program that can fit stellar population synthesis templates to given photometric information. We use FAST to fit models to the 17 galaxies with data in the NSA catalog using the \cite{Bruzual} stellar population synthesis model, the \cite{Calzetti} model for dust attenuation, a \cite{Chabrier} initial mass function (IMF), and an exponential star formation history. We then integrate the best-fit model spectra to derive mass-to-light ratios in the $r$-band and in the IRAC 4.5 $\mu$m band, which we denote $\Upsilon_{*,r}$ and $\Upsilon_{*, 4.5}$, respectively.

We compare the results from FAST against a second method. The program {\tt kcorrect}, as presented in \cite{kcorrect}, is also based on \cite{Bruzual} models and a \cite{Chabrier} initial mass function (IMF). However, instead it fits the photometry as a linear combination of spectral energy distribution model templates. We ran this procedure on the same 17 galaxies. 

\begin{figure*}
\centering
\includegraphics[width=1.0\textwidth]{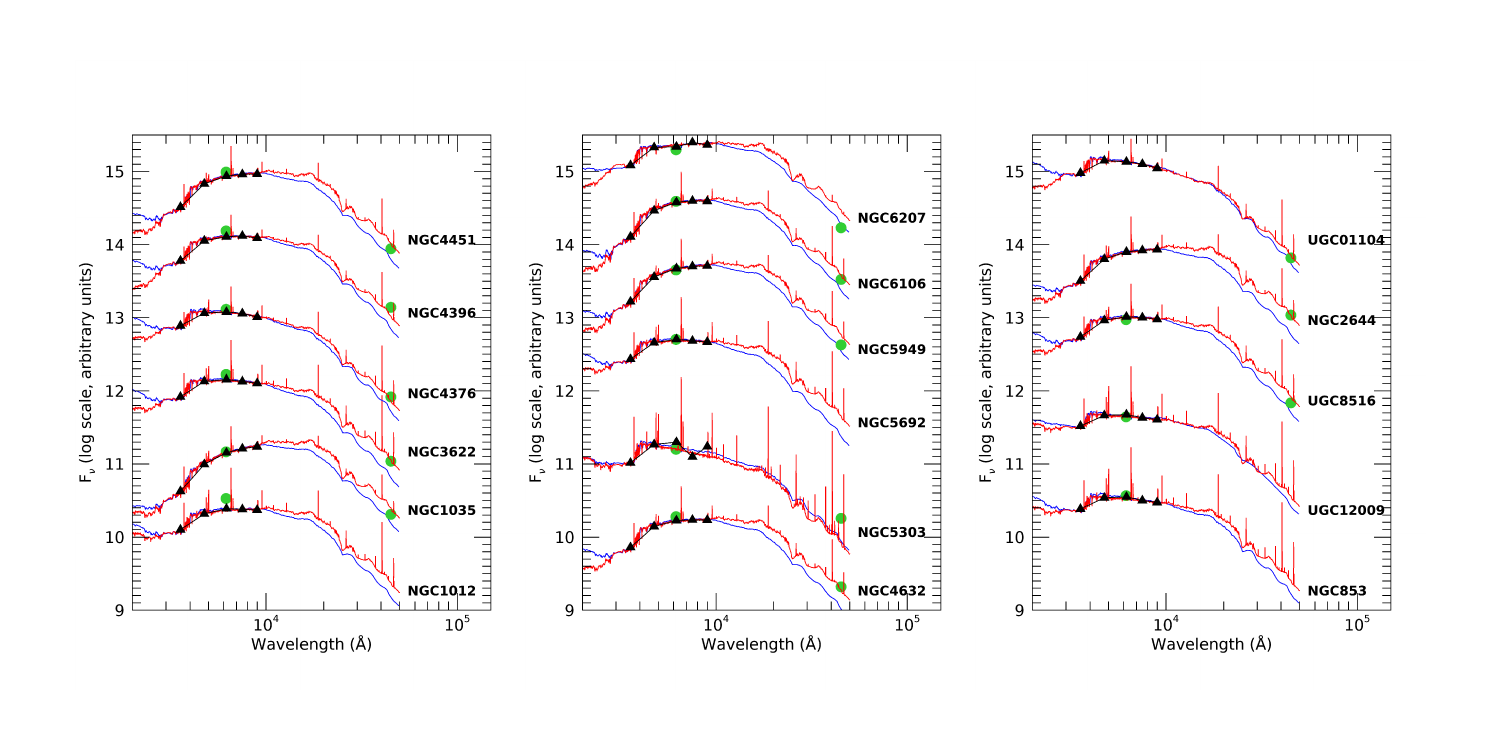}
\caption{Photometry and stellar population synthesis model fits for the 17 galaxies in the NSA catalog. FAST models are represented in blue and {\tt kcorrect} is represented in red. The photometric data used to constrain the models (from the NSA) are black triangles, while the $r$-band and 4.5 $\mu$m luminosities measured from photometry in Section \ref{MGE} are represented by green points. Both models fit the input data well, in addition to the $r$-band luminosities, however the {\tt kcorrect} models are better predictors of the infrared data when available.}
\label{fig:flux}
\end{figure*}

The 4.5 $\mu$m data were not used to constrain the fits and so they provide a test of the models' ability to predict the near-infrared luminosities. The mean difference between the IRAC 4.5 $\mu$m flux and the FAST-predicted flux at the same wavelength is 0.19 dex, compared to a mean difference of -0.02 dex for {\tt kcorrect}. The models from {\tt kcorrect} are both an accurate fit to the input optical data as well as a better predictor of the 4.5 $\mu$m luminosities, and thus we chose to proceed using the values from this method. Comparison plots of the FAST and {\tt kcorrect} models alongside the data can be found in Figure \ref{fig:flux}. 

\begin{figure*}
\centering
\includegraphics[width=0.9\textwidth]{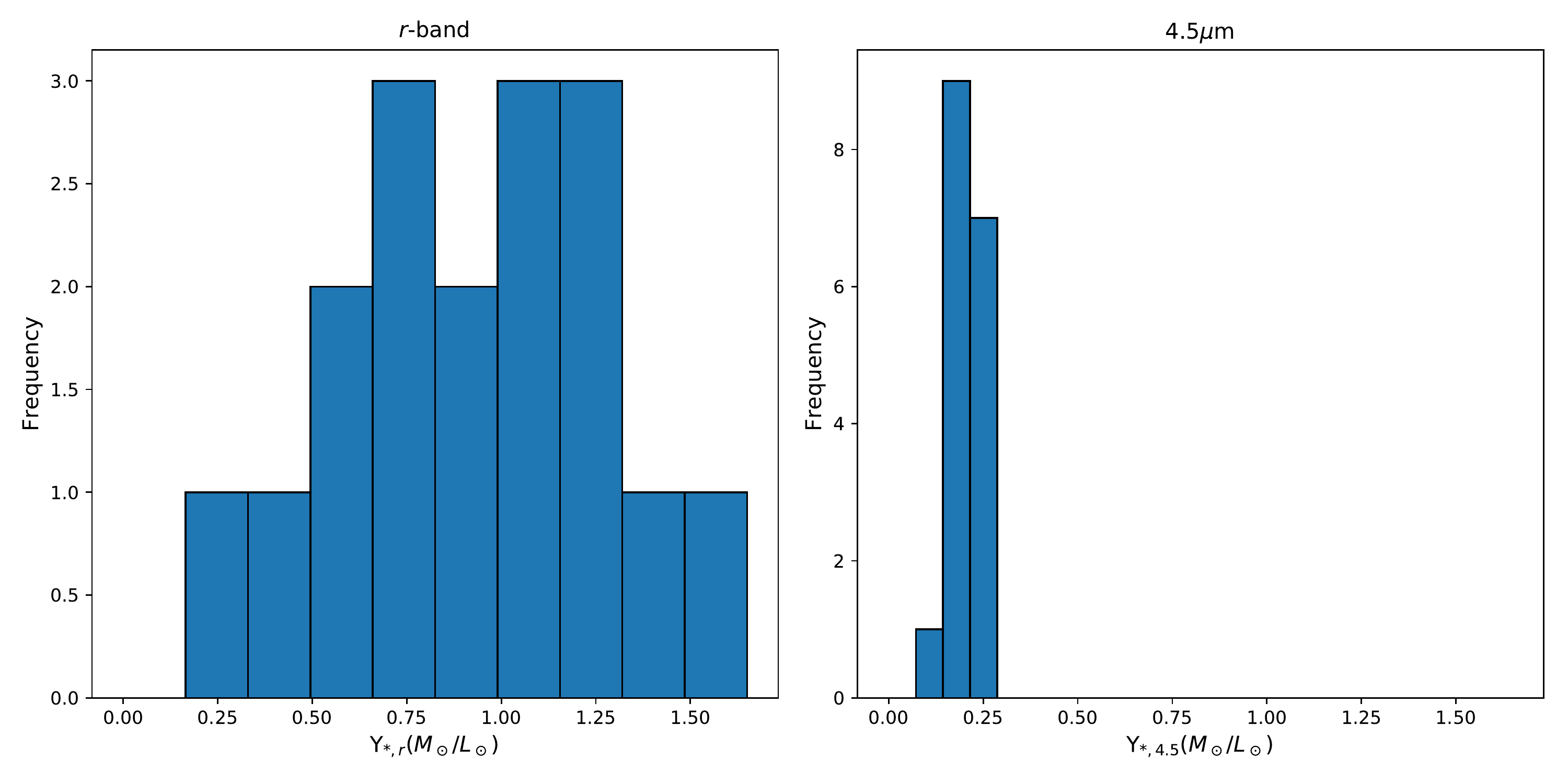}
\caption{Histograms of the {\tt kcorrect} mass-to-light ratios in the $r$-band (left) and 4.5 $\mu$m (right) for the 17 galaxies in the NSA catalog. The close grouping in the 4.5 $\mu$m data led us to adopt a constant value of $\Upsilon_{*,4.5}= 0.21 \msol / \textrm{L}_\sun$ for all galaxies with infrared data.}
\label{fig:MLhist}
\end{figure*}

Histograms of our results from {\tt kcorrect} for each wavelength band can be found in Figure \ref{fig:MLhist}. The mass-to-light values for 4.5 $\mu$m luminosities were closely grouped across our sample (scatter is 20\% of mean), leading us to adopt the mean value of $\Upsilon_{*,4.5} = 0.21 \msol / \textrm{L}_\sun$ for all galaxies with data in this band. The values of $\Upsilon_{*,r}$ however show more variation within the sample (scatter is 37\% of mean). This is consistent with our expectation that the mass-to-light ratio is less sensitive to age, dust, and metallicity in the near-infrared than at optical wavelengths \citep{mcgaugh}.


There is a wide range in SPS model predictions in the near-infrared, as reviewed by \cite{MandS14}. The most important model dependences are the treatment of TP-AGB stars and the assumed IMF. Dynamical constraints are needed to calibrate the absolute mass scale. The DiskMass Survey \citep{Diskmass} consider a sample of 30 spiral galaxies with masses $M_* \sim 10^{10.3} \msol$. They use measurements of the vertical stellar velocity dispersion to find an estimate of $\Upsilon_{*,K}$ that is independent of stellar population synthesis models, finding $\Upsilon_{*,K} = 0.31 \pm 0.07 \msol / \textrm{L}_\sun$. Converting with $\Upsilon_{*,4.5} = 0.91 \times \Upsilon_{*,K} -0.08$ \citep{Oh-ML} gives a value of $\Upsilon_{*,4.5}^{\rm DiskMass} = 0.20 \msol / \textrm{L}_\sun$, in excellent agreement with our SPS-based estimate of $\Upsilon_{*,4.5} = 0.21 \msol / \textrm{L}_\sun$. 

Although this agreement is reassuring, we note that other methods have delivered different results. \cite{MandS15} calibrated the tight relation between rotation speed and baryonic mass (the baryonic Tully-Fisher relation) with gas-rich galaxies; they then applied this relation to normal galaxies to estimate $\Upsilon_*$ and inferred a typical value about twice that of the DiskMass Survey. They conclude that, at present, there is a factor of two systematic
uncertainty in the mass-to-light ratio. We will explore a wide range of $\Upsilon_*$ in Section \ref{ML_robust} to evaluate the effect of this uncertainty on our dark matter profile measurements.

\begin{figure}
\centering
\includegraphics[width=\columnwidth]{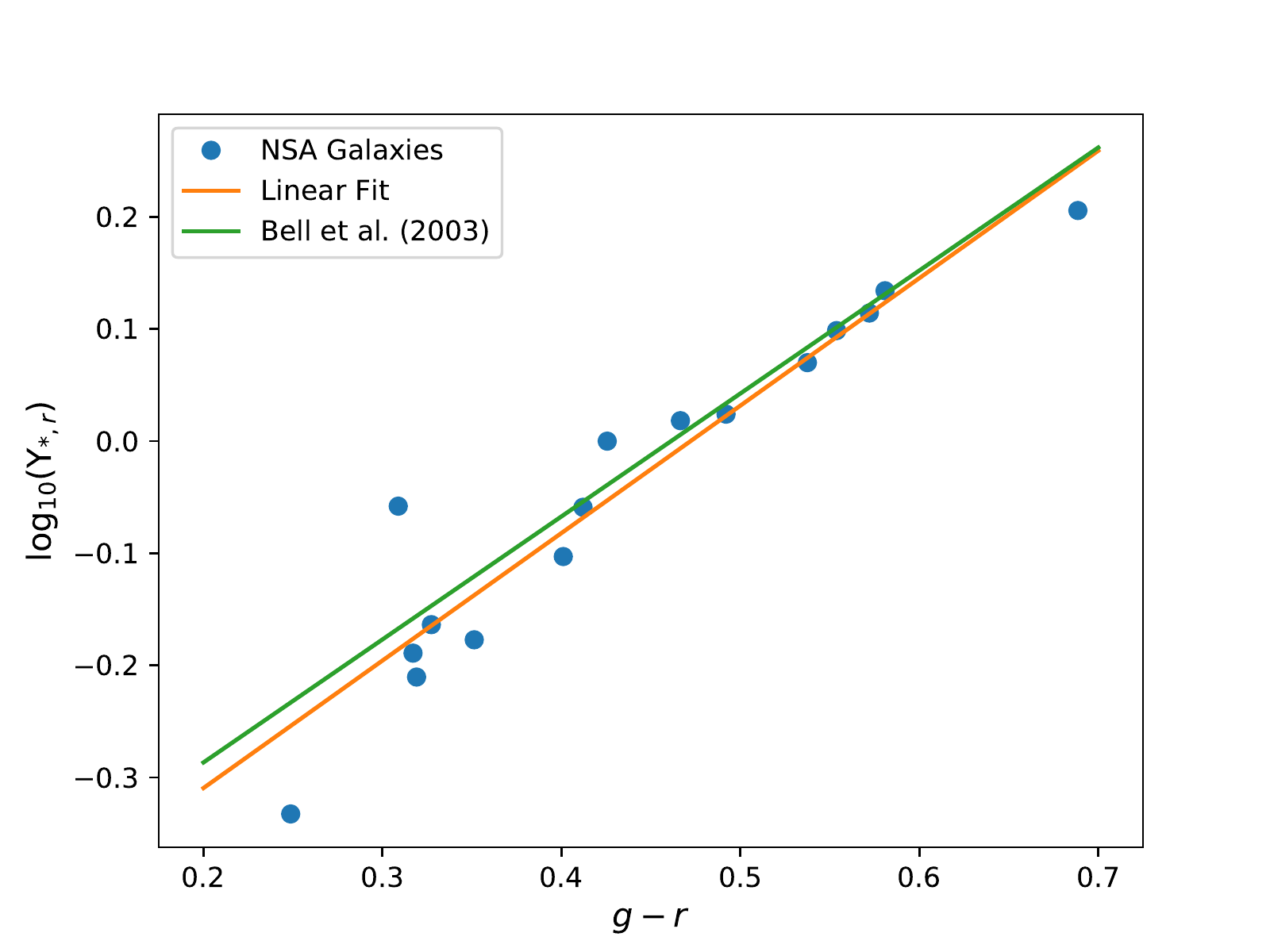} 
\caption{The blue dots represent the galaxies in our sample that are found in the NSA catalog. The values of $\Upsilon_{*,r}$ are obtained using {\tt kcorrect} and the $g-r$ colors are taken from the catalog. The linear fit (orange), given by $\log_{10}(\Upsilon_{*,r}) = 1.14(g-r) -0.54$, has a scatter of 0.17$\msol / \textrm{L}_\sun$. The green line is reproduced from \cite{Bell}.} 
\label{fig:gr-plot}
\end{figure}

While a single value of $\Upsilon_{*,4.5}$ suffices for the infrared data, we find a much larger scatter in the $r$-band (see Figure \ref{fig:MLhist}). This presents a problem for estimating $\Upsilon_{*,r}$ for the 9 galaxies in our sample that are not present in the NSA catalog. The $g-r$ color has been shown to correlate well with the optical mass-to-light ratio (e.g., \citealt{Bell}), which motivates us to explore this correlation using the 17 NSA galaxies. We plotted their $g-r$ magnitudes against the $r$-band mass-to-light ratios found with {\tt kcorrect}, as shown in Figure \ref{fig:gr-plot}, and we fit a line to the data, enabling us to predict the value of $\Upsilon_{*,r}$ using only the $g-r$ color of a galaxy. 

The fit has a linear scatter of $0.17 \msol / \textrm{L}_\sun$. \cite{Bell} provides a relation between mass-to-light ratio and color, finding $\log_{10}\Upsilon_{*,r} = 1.14(g-r) -0.54$. As a comparison, we plot this line alongside our linear fit in Figure \ref{fig:gr-plot} and find excellent agreement between the two. For consistency, we will use the mass-to-light ratio predicted by our linear fit for all galaxies. 

In order to predict $\Upsilon_{*,r}$ for the galaxies not in the NSA, we require $r$-band magnitudes, which we already have for all galaxies in our sample (see Paper 1), and $g$-band magnitudes. Of the galaxies not in the NSA, five have $g$-band imaging taken with SPICAM on the 3.5m telescope at Apache Point Observatory on 2013 October 06 and 2014 November 11. We use the MGE process described in Section \ref{MGE} to determine the $g$-band magnitudes. The $g$-band magnitude for NGC 2976 is provided by the Spitzer Local Volume Legacy Survey \citep{LVL}. Three galaxies did not have $g$-band data available, so we instead calculate the $B-V$ color and convert it to $g-r$ color following \cite{B-V}. NGC 6503 has $B$- and $V$-band data from \cite{LVL}, while NGC 7320 and UGC 3371 have the same from \cite{third}.

\begin{deluxetable}{lcccc} 
\tablecolumns{5}
\tablewidth{0pt}
\tablecaption{Stellar Mass-to-Light Ratio Estimates}
\tablehead{\colhead{Galaxy} & \colhead{$\Upsilon_* \pm \sigma$} & \colhead{$\log M_*$} & \colhead{$\Upsilon_*$} & \colhead{$\lambda$} \\ \colhead{} & \colhead{(SPS est.)} & \colhead{($\msol$)} & \colhead{(Max. Disk.)} & \colhead{}}
\startdata
NGC 746 & 0.87 $\pm$ 0.15 & 8.9 & 0.67 & $r$-band \\ 
NGC 853 & 0.67 $\pm$ 0.15 & 9.3 & 1.59 & $r$-band \\ 
NGC 949 & 1.09 $\pm$ 0.15 & 9.3 & 1.35 & $r$-band \\ 
NGC 959 & 0.21 $\pm$ 0.04 & 8.8 & 0.53 & 4.5 $\mu$m \\ 
NGC 1012 & 0.99 $\pm$ 0.15 & 9.4 & 0.18 & $r$-band \\ 
NGC 1035 & 0.21 $\pm$ 0.04 & 9.6 & 0.22 & 4.5 $\mu$m \\ 
NGC 2644 & 0.21 $\pm$ 0.04 & 9.7 & 0.47 & 4.5 $\mu$m \\ 
NGC 2976 & 0.21 $\pm$ 0.04 & 8.9 & 0.28 & 4.5 $\mu$m \\ 
NGC 3622 & 0.21 $\pm$ 0.04 & 9.2 & 0.06 & 4.5 $\mu$m \\ 
NGC 4376 & 0.21 $\pm$ 0.04 & 9.1 & 0.53 & 4.5 $\mu$m \\ 
NGC 4396 & 0.21 $\pm$ 0.04 & 9.2 & 0.42 & 4.5 $\mu$m \\ 
NGC 4451 & 0.21 $\pm$ 0.04 & 9.6 & 0.27 & 4.5 $\mu$m \\ 
NGC 4632 & 0.21 $\pm$ 0.04 & 9.5 & 0.41 & 4.5 $\mu$m \\ 
NGC 5303 & 0.21 $\pm$ 0.04 & 9.7 & 0.24 & 4.5 $\mu$m \\ 
NGC 5692 & 0.83 $\pm$ 0.15 & 9.5 & 1.06 & $r$-band \\ 
NGC 5949 & 0.21 $\pm$ 0.04 & 9.3 & 0.50 & 4.5 $\mu$m \\ 
NGC 6106 & 0.21 $\pm$ 0.04 & 9.7 & 0.22 & 4.5 $\mu$m \\ 
NGC 6207 & 0.21 $\pm$ 0.04 & 9.6 & 0.31 & 4.5 $\mu$m \\ 
NGC 6503 & 0.21 $\pm$ 0.04 & 9.4 & 0.22 & 4.5 $\mu$m \\ 
NGC 7320 & 0.21 $\pm$ 0.04 & 9.0 & 0.53 & 4.5 $\mu$m \\ 
UGC 1104 & 0.21 $\pm$ 0.04 & 8.1 & 0.35 & 4.5 $\mu$m \\ 
UGC 3371 & 0.41 $\pm$ 0.15 & 8.5 & 0.89 & $r$-band \\ 
UGC 4169 & 0.21 $\pm$ 0.04 & 9.4 & 0.26 & 4.5 $\mu$m \\ 
UGC 8516 & 0.21 $\pm$ 0.04 & 9.1 & 0.14 & 4.5 $\mu$m \\ 
UGC 11891 & 0.62 $\pm$ 0.15 & 8.9 & 1.55 & $r$-band \\ 
UGC 12009 & 0.67 $\pm$ 0.15 & 9.1 & 0.45 & $r$-band 
\enddata
\tablecomments{Stellar masses are calculated using the SPS-predicted value of $\Upsilon_*$. The maximum disk $\Upsilon_*$ is found through scaling the stellar contribution to the rotation curve, see Section \ref{minmax} for more detail. } \label{ML_table}
\end{deluxetable}
Table \ref{ML_table} gives the values of $\Upsilon_*$ as well as the wavelength used for mass models ($r$-band or 4.5 $\mu$m).

\subsection{Maximum and Minimum Disk Estimates of $\Upsilon_*$}\label{minmax}
In order to test the sensitivity of our derived dark matter profiles to our estimates of the stellar distribution, we wish to explore the full range of plausible $\Upsilon_*$. In addition to the estimates based on SPS models described in the previous section, we will also consider limits on $\Upsilon_*$ that are based only on kinematic constraints. The ``minimum disk'' and ``maximum disk'' hypotheses bracket the kinematically allowed values of $\Upsilon_*$. 

The minimum value the mass-to-light ratio can take is when there is hypothetically no contribution to the gravitational potential from the stars. In this case, $\Upsilon_* = 0$ and the rotation curve is a dark matter-only model (we will also assume gas is negligible, see Section \ref{gas} for further discussion). A common procedure for estimating the maximum $\Upsilon_*$ value is to increase the amplitude of the model stellar rotation curve until it exceeds the data at some radius. This simple approach has some shortcomings, however, which necessitates minor modifications. 

First, the radius at which the stellar rotation curve exceeds the data is generally found to be the innermost observed radius, which often leads to an unphysically small estimate of the maximum $\Upsilon_*$. This is likely due to random uncertainties in the rotation curve that this procedure does not account for, as well as any non-circular motions that might affect the innermost point. As a more conservative estimate of the maximum $\Upsilon_*$, we instead consider the value that produces a stellar rotation curve that exceeds the total rotation curve at two radial bins. 

While an improvement over surpassing one radial bin, there remain several special cases we must address. For four galaxies (NGC 959, NGC 4376, NGC 7320, UGC11891), surpassing two radial bins leads instead to an unphysically large value of $\Upsilon_*$, motivating us to set an upper limit of 2.5 times the value of $\Upsilon_*$ predicted from SPS models. There were also seven cases where the maximum disk estimate of $\Upsilon_*$ was smaller than the SPS-predicted value. For four of these cases, studying the maximum disk case is inconsequential because we ultimately find that these galaxies are not suitable for measuring the dark matter profile (see Section \ref{grade}), however three require addressing. The SPS-predicted $\Upsilon_*$ value for UGC 12009 already surpasses two (nearly three) radial bins out of a total of five (as seen in Figure \ref{fig:final_plots}). As the maximum disk value could not reasonably be larger, we keep the original estimate, despite being lower than the SPS-predicted value. For NGC 2976 and NGC 6503, there is a small peak in the inner region of the stellar model rotation curve that causes the low estimate of $\Upsilon_*$ when surpassing two radial bins. We therefore define the maximum disk $\Upsilon_*$ as the value that exceeds the first radial bin that falls outside the range affected by the peak. Such peaks arise from a bright nucleus, which is likely to have a different mass-to-light ratio than the rest of the galaxy, and so their omission from our maximum disk estimate is justified. 

Although these criteria are ad hoc, they are tailored to our goal of studying the inner mass distribution. We found it important to have a maximum disk definition that ensured physical results within $\sim 1$ kpc. For comparison, we also considered the definition of the maximum disk presented by \cite{Sackett}, in which the disk produces 85\% of the velocity at 2.2 times the disk scale radius. This produced a higher maximum disk $\Upsilon_*$ than our previous method in 12 of 18 galaxies, and in almost all of those cases, the stellar contribution would significantly exceed the total rotation throughout all or most of the inner kpc. This may be related to the fact that the \cite{Sackett} maximum disk estimate is based on observations of much more massive galaxies, where the variation of the stellar mass fraction with radius is likely different.

The final values of the maximum disk $\Upsilon_*$ can be found in Table \ref{ML_table}.  We explore the effects of these values on our mass models in Section \ref{ML_robust}. 

\section{Gas Mass Distribution} \label{gas}
Investigations of dark matter in dwarf galaxies around $M_* \sim10^9 \msol$ often neglect the gas mass (e.g., \citealt{Josh05, ghasp}), based on the assumption that it is subdominant to the stars. We test this assumption by including estimates of the contribution to the rotation curves from atomic and molecular gas, and will perform our analysis both with and without the gas included to determine what effects, if any, the inclusion of gas kinematics has on the mass model. Here we describe how we determined the contribution to the rotation curves due to gas, while we examine the resulting mass models in Section \ref{gas_model}. 

\subsection{Atomic Gas}\label{HI}
While most of our galaxies do not have resolved \ion{H}{1} data, total \ion{H}{1} fluxes exist in the literature for all of the galaxies in our sample. We are able to use these total flux values to estimate the \ion{H}{1} mass surface density, which can then be used to estimate the contribution to the rotation curve. 

Some galaxies have only one source giving a value for the \ion{H}{1} flux, but for the cases with multiple sources, we preferentially select those that had been corrected for beam attenuation, pointing offsets, and \ion{H}{1} self-absorption. Data for 13 of our galaxies are found in \cite{Springob}, which includes all of the aforementioned corrections, six are found in \cite{Huchtmeier}, \cite{Doyle} and \cite{Popping} each contain two, while \cite{Schneider}, \cite{Greisen}, and \cite{Walter} contain one each. The \ion{H}{1} fluxes were converted to a total mass using $M_{\rm HI} = (2.36 \times 10^5) D^2 f$, where D is the distance in Mpc, $f$ is the flux in Jy km s$^{-1}$, and $M_{\rm HI}$ is in solar masses \citep{HI_mass}. 

We follow the method in \cite{Diskmass} to use these values to estimate the \ion{H}{1} mass distribution. They find the radial \ion{H}{1} mass surface density profile is well fit with a Gaussian as follows: 
\begin{equation}
\Sigma_{\rm HI}(R) = \Sigma_{\rm HI}^{\rm max} \exp \left[-\frac{(R-R_{\Sigma, \rm max})^2}{2\sigma_{\Sigma}^2}\right]
\end{equation}
where $\Sigma_{\rm HI}^{\rm max}$ is the peak density, $R_{\Sigma,\rm max}$ is the radius at which the peak occurs, and $\sigma_{\Sigma}$ is the width of the profile. Since $\Sigma_{\rm HI}^{\rm max}$ is a normalization constant, it is determined explicitly by integrating over the disk. The remaining values are found using $R_{\rm HI}$, defined as the radius for which $\Sigma_{\rm HI} = 1 \msol \mathrm{pc} ^{-2}$. There is a relationship between $R_{\rm HI}$ and the total \ion{H}{1} mass, though it varies somewhat depending on the mass range of the sample. The galaxies used in \cite{Diskmass} are more massive than ours, so we use the relation from \cite{Swaters} as their mass range is closer to ours. They find:
\begin{equation}
\log M_{\rm HI}= 1.86 \log(2 R_{\rm HI}) + 6.6,
\end{equation}
where $M_{\rm HI}$ is measured in units of $\msol$ and $R_{\rm HI}$ is in kpc. Taking the radial \ion{H}{1} surface density profile (averaged over all of the dwarf galaxies in the \citealt{Swaters} sample), we fit a Gaussian to determine $R_{\Sigma,\rm max} = 0.20 R_{\rm HI}$ and $\sigma_{\Sigma} = 0.44 R_{\rm HI}$. Finally, we multiply $\Sigma_{\rm HI}$ by 1.4 to account for helium and metals, $\Sigma_{\rm atomic}  = 1.4\Sigma_{\rm HI}$ \citep{Diskmass}. The values used in this calculation for each galaxy can be found in Table \ref{HI-table}.

The \ion{H}{1} surface densities are then used to estimate the atomic gas contribution to the rotation curve. This is done analytically, following the method in Section 2.6.3 of \cite{galactic}, which derives an expression for $v_{\rm circ}$ for a thin disk given the surface density $\Sigma(R)$. 


\begin{deluxetable}{lccc} 
\tablecolumns{4}
\tablewidth{0pt}
\tablecaption{\ion{H}{1} Parameters}
\tablehead{\colhead{Galaxy} & \colhead{\ion{H}{1} Flux} & \colhead{$M_{\rm HI}$} & \colhead{$R_{\rm HI}$} \\
\colhead{} & \colhead{(Jy km s$^{-1}$)} & \colhead{($10^8 \msol$)}  & \colhead{(kpc)} }
\startdata
NGC 746 & 18.6\phn & 6.3& 7.6 \\
NGC 853 & 6.0 & 5.7& 7.2 \\ 
NGC 949 & 16.4\phn & 3.9 & 5.9 \\ 
NGC 959 & 16.7\phn & 4.0 & 5.9 \\ 
NGC 1012 & 55.1\phn & 25.5\phn & 16.1\phn \\ 
NGC 1035 & 14.2\phn & 8.6 & 9.0 \\ 
NGC 2644 & 3.7 & 7.9 & 8.6 \\ 
NGC 2976 & 45.4\phn & 1.4 & 3.4 \\ 
NGC 3622 & 13.0\phn & 16.2\phn & 12.7\phn \\ 
NGC 4376 & 3.7 & 5.0 & 6.7 \\ 
NGC 4396 & 21.4\phn & 12.9\phn & 11.2\phn \\ 
NGC 4451 & 3.2 & 5.0 & 6.8 \\ 
NGC 4632 & 55.9\phn & 25.8\phn & 16.2\phn \\ 
NGC 5303 & 11.3\phn & 16.7\phn & 12.8\phn \\ 
NGC 5692 & 3.8 & 6.5 & 7.7 \\ 
NGC 5949 & 5.8 & 2.3 & 4.4 \\ 
NGC 6106 & 33.0\phn & 44.9\phn & 21.9\phn \\ 
NGC 6207 & 34.4\phn & 20.8\phn & 14.5\phn \\ 
NGC 6503 & 205.0\phn & 19.2\phn & 13.9\phn \\ 
NGC 7320 & 8.3 & 3.8 & 5.8 \\ 
UGC 1104 & 9.8 & 2.3 & 4.4 \\ 
UGC 3371 & 31.6\phn & 16.8\phn & 12.9\phn \\ 
UGC 4169 & 29.0\phn & 61.7\phn & 25.9\phn \\ 
UGC 8516 & 3.9 & 4.4 & 6.3 \\ 
UGC 11891 & 88.5\phn & 12.4\phn & 10.9\phn \\ 
UGC 12009 & 9.8 & 9.2 & 9.3 
\enddata
\tablecomments{See Section \ref{HI} for derivation details. }\label{HI-table}
\end{deluxetable}

\subsection{Molecular Gas}
We have resolved CO data for the 11 galaxies in our sample that were studied in \cite{Mai1}, which allows us to determine the distribution of molecular gas in these galaxies. Using the CO moment-0 maps provided by \cite{Mai1}, we convert to H$_2$ mass surface density maps using:
\begin{multline}
\bigg[\frac{\Sigma_{\rm H_2}}{\msol \textrm{pc}^{-2}}\bigg] = 1.6 \bigg[\frac{I_{\rm CO}\Delta V}{\textrm{K km s}^{-1}}\bigg]\\
\times\bigg[\frac{X_{\rm CO}}{10^{20} \textrm{cm}^{-2} (\textrm{K km s}^{-1})^{-1}}\bigg] \cos i
\end{multline}
where $i$ is the inclination of the galaxy and $X_{\rm CO}$ is the CO to H$_2$ conversion factor. We use a value of $X_{\rm CO} = 2.0 \times10^{20} \textrm{cm}^{-2} (\textrm{K km s}^{-1})^{-1}$, following the recommendation in \cite{Bolatto}. The resulting H$_2$ mass surface density maps are then multiplied by 1.4 to account for helium and metals. We average the maps over a series of elliptical bins to produce a radial molecular mass surface density profile for each galaxy. 

We again analytically derive the rotation curves by assuming a thin disk and following Section 2.6.3 of \cite{galactic}. Before we can include them in our mass model, we first need to extend the profiles, as several of the CO maps do not extend as far out as our H$\alpha$ data. We do this by fitting an exponential model which is then scaled to agree with the outermost point of the CO data to ensure continuity. When necessary, we use this function to extrapolate the curve to match the extent of the H$\alpha$ data. This did not produce changes in the rotation curves within the inner 1 kpc for any of our galaxies, so this extrapolation does not have a significant effect on our results.

\section{Mass Modeling}
In this section, we describe the procedure for modeling the rotation curves in terms of their baryonic and dark matter components. In addition to discussing the definition of the inner dark matter density slope that is most robustly constrained by our data, we also assess the fit quality of the mass model for each galaxy to determine the subset most suitable for robust measurement of the dark matter profiles. 

\subsection{Modeling the Dark Matter Distribution}\label{MCMC}
The rotation curves derived in Paper 1 represent the combined gravitational force from all of the mass components of the galaxy (dark matter, stars, gas). In order to determine the distribution of dark matter, the various contributions to the rotation curve need to be separated. To do this we use a Markov Chain Monte Carlo (MCMC) fitting procedure to model the rotation curves as a sum of the stellar, gaseous, and dark matter components: $V_{\rm tot}^2 = V_{\rm stars}^2 + V_{\rm gas}^2 + V_{\rm DM}^2$. We use the estimates of the stellar and gaseous rotation curves derived in Section \ref{stars} and \ref{gas} respectively, and model the dark matter as a spherical halo following a generalized Navarro-Frenk-White (gNFW) profile:
\begin{equation}\label{eq:gNFW}
\rho(r) = \frac{\rho_0}{(\frac{r}{r_s})^\beta(1+\frac{r}{r_s})^{3-\beta}}.
\end{equation}
The gNFW profile includes the NFW scale radius $r_s$ and characteristic density $\rho_0$, as well as an additional parameter $\beta$, that dictates the slope as $r \rightarrow 0$ (note that setting $\beta=1$ returns the standard NFW profile). Typically, galaxies are described using a parametrization defined by the virial mass $M_{200}$ (given by the enclosed mass at $r_{200}$, the radius at which the mean enclosed density is equal to 200 times the critical density) and the concentration $c_{200} = r_{200}/r_s$. In the case of the generalized NFW profile, we found that parameter covariance was reduced by recasting Equation \ref{eq:gNFW} in terms of the parameter $r_{-2} = (2-\beta) r_s$, which is the radius at which $\rho \propto r^{-2}$ locally (see  e.g., \citealt{Drew}; note that $r_{-2}$ reduces to $r_s$ for the NFW case where $\beta = 1$), and the concentration $c_{-2} = r_{200} / r_{-2}$. 

In addition to the three parameters describing the gNFW model, we also include the stellar mass-to-light ratio $\Upsilon_*$ as a parameter with a prior based on the results of our population synthesis analysis from Section \ref{ML}. Assuming Gaussian errors, the likelihood function $L$, including the full covariance matrix, is given by 
\begin{equation}
\log L = -N \log j - \frac{1}{2}(V_{t} - m)^T\bullet (\frac{C^{-1}}{j^2})\bullet (V_{t} - m),\label{gnfw-eq}
\end{equation}
where $\bullet$ represents matrix multiplication, $N$ is the number of points in the rotation curve data given by $V_t$, $m$ represents the model of the total rotation curve (dark matter, stars, and gas), $C$ is the $N \times N$ covariance matrix (obtained from the DiskFit output as described in Paper 1),  and $j$ represents a jitter term. In order to control the errors, we included this ``jitter" term, which is meant to aid in the cases where the model does not describe the data well by increasing the errors until a reasonable fit is obtained (see e.g., \citealt{jitter}). This means $j \sim 1$ when the model adequately describes the data within the uncertainties, while $ j > 1$ indicates the model is inadequate or the uncertainties are underestimated. We also use the full covariance matrix from the bootstrapping analysis in order to take into account the correlated uncertainties among radial bins (see Paper 1 for details).  

We use the MCMC ensemble sampler {\tt emcee} \citep{emcee}, setting uniform priors of $0.0 < \beta < 2.0, 8 < \log M_{200} < 12.0, 1.0 < c_{-2} < 40.0$ and Gaussian priors on $\log \Upsilon_*$ and $\log j$. The prior for $\Upsilon_*$ is centered on the SPS value with a width dictated by the uncertainty (see Table \ref{ML_table}), while $\log j$ is centered on 0 with a dispersion of 0.7, corresponding to a factor of 2 in $j$. We first initialize the {\tt emcee} walkers at positions drawn from these priors then use a short initial run (500 steps) to estimate rough parameter values. These estimates are then used to initialize the final MCMC run of 50 walkers and 10,000 steps. We use only the second half of the chains in our analysis.

\begin{deluxetable*}{lcccccc} 
\tablecolumns{7}
\tablewidth{0pt}
\tablecaption{Inner Dark Matter Density Profile Slope}
\tablehead{\colhead{Galaxy} & \colhead{$\beta^* \pm \sigma$} & \colhead{$\beta^* \pm \sigma$} & \colhead{$\beta^* \pm \sigma$} & \colhead{$\log j \pm \sigma$} &\colhead{Grade} \\ \colhead{} & \colhead{} & \colhead{(Min. Disk)} & \colhead{(Max. Disk)} & \colhead{} & \colhead{}}
\startdata
NGC 959 & 1.25 $\pm$ 0.12 & 1.21 $\pm$ 0.09 & 1.34 $\pm$ 0.24 & 0.28 $\pm$ 0.19 & 1 \\ 
NGC 1035 & 0.58 $\pm$ 0.11 & 0.95 $\pm$ 0.21 & 0.51 $\pm$ 0.10 & 0.50 $\pm$ 0.16 & 1 \\ 
NGC 2976 & 0.33 $\pm$ 0.05 & 0.42 $\pm$ 0.03 & 0.31 $\pm$ 0.08 & -0.07 $\pm$ 0.14 & 1 \\ 
NGC 4376 & 0.70 $\pm$ 0.10 & 0.70 $\pm$ 0.08 & 0.68 $\pm$ 0.15 & 0.37 $\pm$ 0.22 & 1 \\ 
NGC 4396 & 0.84 $\pm$ 0.12 & 0.86 $\pm$ 0.10 & 0.81 $\pm$ 0.14 & -0.15 $\pm$ 0.16 & 1 \\ 
NGC 5303 & 0.54 $\pm$ 0.19 & 0.91 $\pm$ 0.11 & 0.50 $\pm$ 0.25 & 0.21 $\pm$ 0.26 & 1 \\ 
NGC 5692 & 0.76 $\pm$ 0.24 & 1.09 $\pm$ 0.10 & 0.74 $\pm$ 0.46 & 0.35 $\pm$ 0.23 & 1 \\ 
NGC 5949 & 0.84 $\pm$ 0.08 & 0.83 $\pm$ 0.04 & 1.11 $\pm$ 0.32 & 0.47 $\pm$ 0.21 & 1 \\ 
NGC 6207 & 0.73 $\pm$ 0.13 & 0.79 $\pm$ 0.07 & 0.60 $\pm$ 0.14 & 0.35 $\pm$ 0.14 & 1 \\ 
NGC 7320 & 1.03 $\pm$ 0.07 & 1.04 $\pm$ 0.16 & 1.05 $\pm$ 0.11 & 0.18 $\pm$ 0.17 & 1 \\ 
UGC 4169 & 0.59 $\pm$ 0.16 & 0.58 $\pm$ 0.13 & 0.59 $\pm$ 0.17 & 0.36 $\pm$ 0.17 & 1 \\ 
 & & & & & & \\
NGC 853 & 1.10 $\pm$ 0.25 & 0.95 $\pm$ 0.16 & 1.58 $\pm$ 0.45 & 0.94 $\pm$ 0.26 & 2 \\ 
NGC 2644 & 0.70 $\pm$ 0.23 & 0.69 $\pm$ 0.16 & 1.14 $\pm$ 0.54 & 0.02 $\pm$ 0.22 & 2 \\ 
NGC 6106 & 0.59 $\pm$ 0.14 & 0.68 $\pm$ 0.06 & 0.58 $\pm$ 0.15 & 0.57 $\pm$ 0.15 & 2 \\ 
NGC 6503 & 0.72 $\pm$ 0.04 & 0.81 $\pm$ 0.03 & 0.70 $\pm$ 0.04 & 0.70 $\pm$ 0.12 & 2 \\ 
UGC 1104 & 0.67 $\pm$ 0.35 & 1.21 $\pm$ 0.38 & 0.84 $\pm$ 0.46 & 0.41 $\pm$ 0.33 & 2 \\ 
UGC 11891 & 1.00 $\pm$ 0.31 & 0.95 $\pm$ 0.28 & 1.01 $\pm$ 0.34 & 0.07 $\pm$ 0.20 & 2 \\ 
UGC 12009 & 1.20 $\pm$ 0.54 & 1.26 $\pm$ 0.19 & 1.09 $\pm$ 0.50 & 0.19 $\pm$ 0.34 & 2 \\ 
 & & & & & & \\
NGC 746 & 0.64 $\pm$ 0.32 & 0.67 $\pm$ 0.15 & 0.58 $\pm$ 0.23 & 0.24 $\pm$ 0.18 & 3 \\ 
NGC 949 & 1.60 $\pm$ 0.42 & 1.41 $\pm$ 0.11 & 1.59 $\pm$ 0.46 & 0.73 $\pm$ 0.19 & 3 \\ 
NGC 1012 & 1.09 $\pm$ 0.62 & 0.40 $\pm$ 0.09 & 0.30 $\pm$ 0.09 & 1.26 $\pm$ 0.18 & 3 \\ 
NGC 3622 & 0.76 $\pm$ 0.38 & 0.70 $\pm$ 0.18 & 0.67 $\pm$ 0.21 & 0.44 $\pm$ 0.26 & 3 \\ 
NGC 4451 & 1.48 $\pm$ 0.48 & 0.93 $\pm$ 0.22 & 1.52 $\pm$ 0.49 & 0.66 $\pm$ 0.25 & 3 \\ 
NGC 4632 & 1.31 $\pm$ 0.15 & 1.33 $\pm$ 0.08 & 1.21 $\pm$ 0.47 & 0.92 $\pm$ 0.14 & 3 \\ 
UGC 3371 & 0.47 $\pm$ 0.14 & 0.42 $\pm$ 0.13 & 0.50 $\pm$ 0.16 & 0.69 $\pm$ 0.22 & 3 \\ 
UGC 8516 & 0.84 $\pm$ 0.56 & 0.38 $\pm$ 0.12 & 0.47 $\pm$ 0.28 & 1.10 $\pm$ 0.24 & 3 
\enddata
\tablecomments{See Section \ref{beta*} for the definition of $\beta^*$ and Section \ref{MCMC} for details about $\log j$ \label{beta_table}}
\end{deluxetable*}

\subsection{Fit Quality and Suitability of Galaxies for Dark Matter Measurement}\label{grade}
\begin{figure*}
\centering
\includegraphics[width=\textwidth]{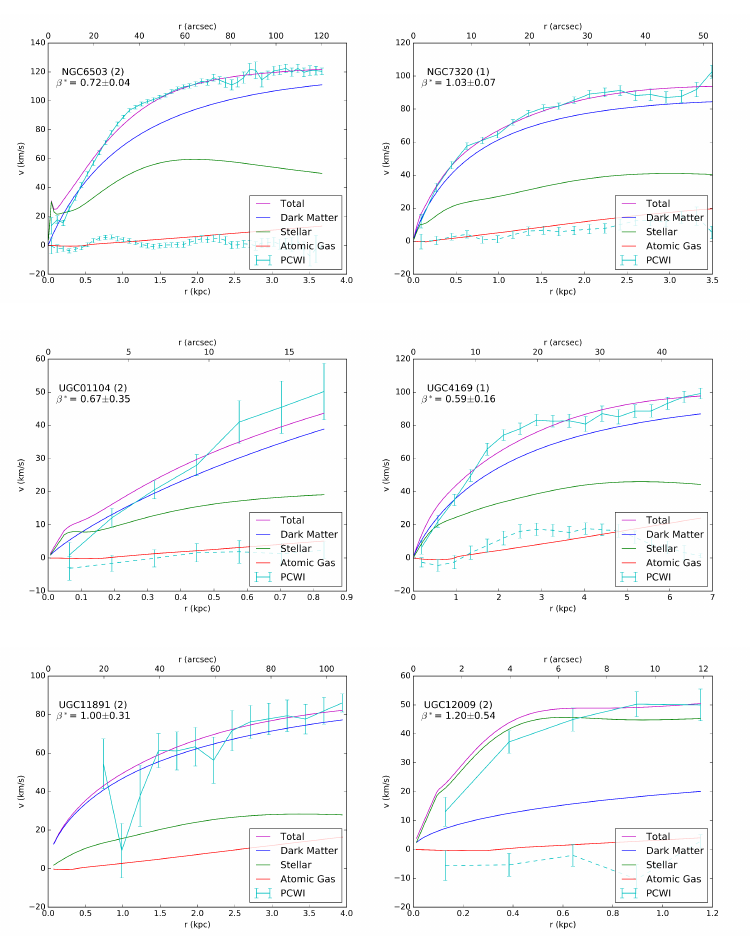}
\caption{Mass models for the galaxies that were graded 1 and 2 (grade 3 galaxies can be found in the Appendix). Each panel contains the galaxy's rotation curve plotted in cyan, derived using data from PCWI. The radial motions are plotted with a dashed line; they are not used in the mass model but are considered in the grading process. We were unable to infer radial motions for two galaxies (UGC 3371 and UGC 11891; see Paper 1 for details). Overlaid are the various components of the mass model (total, dark matter, stars, gas). The galaxy name, grade (in parenthesis), and value of the inner dark matter slope $\beta^*$ (see Section \ref{beta*}) are listed in the upper left of each plot. \label{fig:final_plots}}
\end{figure*}

The MCMC procedure gives us a fiducial mass model for each galaxy that breaks down the contributions from the various mass components to the total rotation curve. Figure \ref{fig:final_plots} shows the model of the PCWI rotation curves (comprising all components) as well as the decomposition into mass component contributions (dark matter, stars, atomic gas, molecular gas when available). While we generally find that our model of the total rotation curve is able to match the PCWI data, not all of our galaxies were fit well, as is to be expected. 

In order to address the potential inadequacies of the model, we assign a grade from 1-3 to each galaxy. We take into account how well the total mass model fit the rotation curve, particularly the slope in the inner parts. We also note if there are strong radial motions, which are plotted in Figure \ref{fig:final_plots} as well. If the radial velocities are a significant fraction of the tangential velocities, it could indicate the presence of irregular or non-circular motions, which may affect our ability to model such galaxies accurately. Finally, we consider the 2D residuals from the DiskFit model. 

As expected, there is a broad correspondence between our subjective grades and the quantitative value of the jitter term $j$. The galaxies with the largest jitter values received poor grades, while the grade 1 galaxies all have $\log j \leq 0.5$. This correspondence reinforces the validity of our subjective grades, which we prefer to use as the deciding criteria as the jitter term only quantifies the goodness of fit of the rotation curve and does not account for other factors, such as the fit quality of the velocity field or evidence of non-circular motions. 


Galaxies assigned grades of 1 or 2 are fit sufficiently well to proceed with further analysis. The eight galaxies assigned a grade of 3 had mass models that we judge did not adequately describe the PCWI data and therefore cannot constrain the value of the inner dark matter density profile. We therefore exclude the grade 3 galaxies for the remainder of the paper. Moving forward only with grades 1 and 2 leaves us a final sample size of 18 galaxies. The grades of individual galaxies can be found in Table \ref{beta_table}. 

The mass models of grade 1 and 2 galaxies are presented in Figure \ref{fig:final_plots}, while the grade 3 galaxies are shown in Figure \ref{fig:final_plots_3}.

\subsection{Definition of the Inner Dark Matter Slope}\label{beta*}
Although the value of $\beta$ in the gNFW profile gives the slope of the dark matter density profile as $r \rightarrow 0$, this is not the best measured quantity as it is naturally asymptotic. Additionally, $\beta$ is often covariant with one or more of the other parameters. Since many of our rotation curves do not fully reach the ``flattened" part, we often cannot constrain all of the gNFW parameters well in our MCMC analysis. In particular, covariance with the scale radius $r_s$ means the data can often be fit with an arbitrarily small $\beta$ and small $r_s$, leading us only to an upper bound on $\beta$. 

\begin{figure*}
\centering
\includegraphics[width=\textwidth]{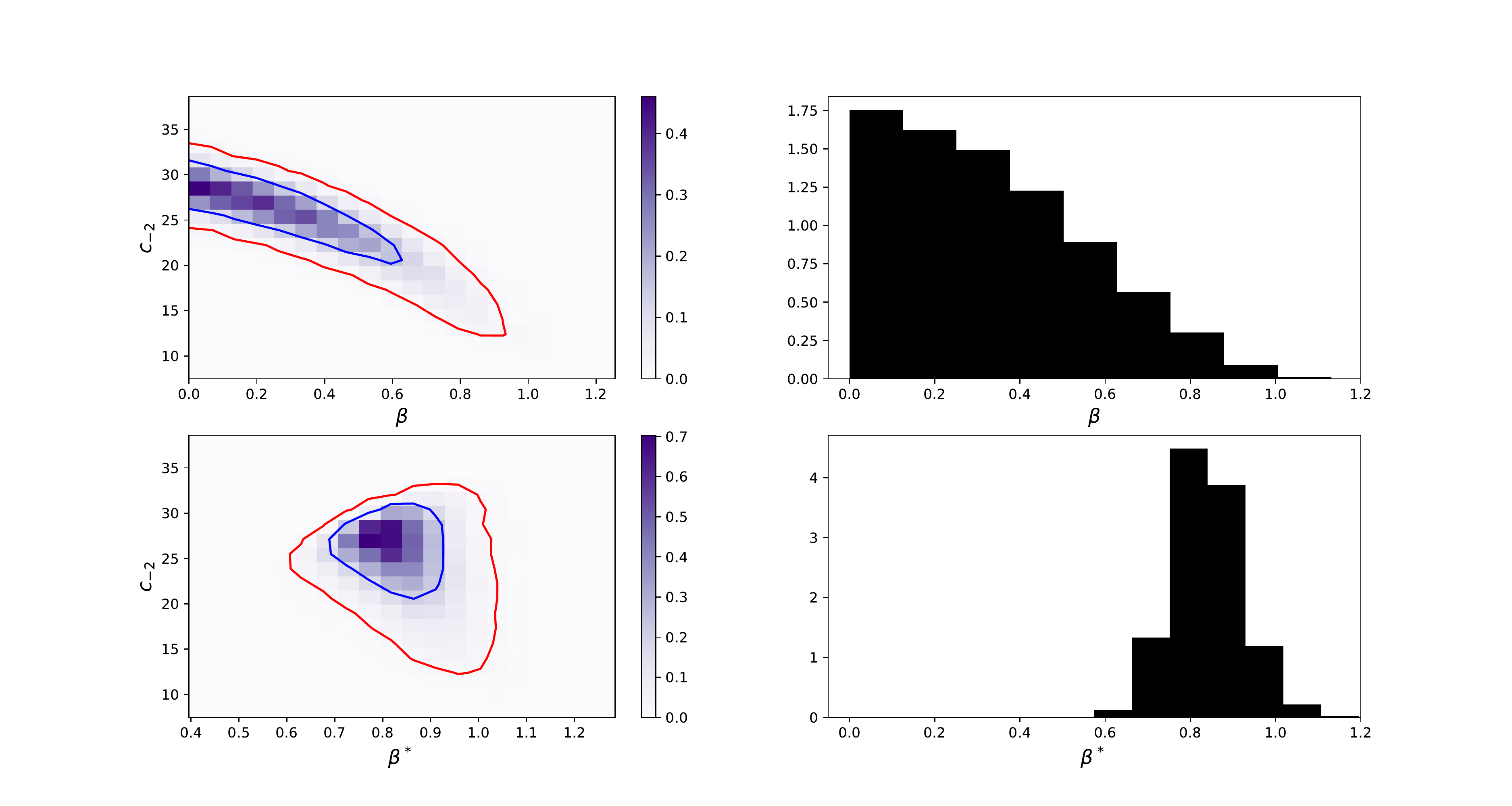} 
\caption{Covariance between concentration and the inner density profile slope of NGC 5949. Left plots show the 2D posterior probability densities for concentration $c_{-2}$ and $\beta$ (top) or $\beta^*$ (bottom) for NGC 5949. The blue and red contours enclose 68\% and 99\% of the posterior respectively. Right plots are marginalized posteriors of each parameter. This motivates our use of $\beta^*$ over $\beta$, as it is clear that $\beta^*$ has less covariance with concentration than $\beta$, and we are able to better constrain the value of $\beta^*$, whereas we can only give an upper bound for $\beta$. }
\label{fig:beta_corner}
\end{figure*}

To remedy this, we decided to define a mean value of the inner slope over a fixed radial range, $\beta^*$. Motivated by the resolution of our PCWI data and the CO data from \cite{Mai1}, as well as by the resolution achieved by simulations examining similar mass galaxies, we define $\beta^*$ over the range 300-800 pc. Taking the posteriors from the analysis described in Section \ref{MCMC}, we calculate the average slope over this range as:
\begin{equation}
\beta^* = -\frac{\log\frac{\rho(0.8 \text{ kpc})}{\rho(0.3 \text{ kpc})}}{\log\frac{0.8}{0.3}}.
\label{betastar}
\end{equation}
We find that $\beta^*$ is much better constrained by the data than the asymptotic $\beta$ from the gNFW profile. As seen in Figure \ref{fig:beta_corner}, there is much less covariance between concentration $c_{-2}$ and $\beta^*$ than with $\beta$, and we are not restricted to citing only an upper bound, instead deriving a tight constraint on $\beta^*$. 
\begin{figure*}
\centering
\includegraphics[width=\textwidth]{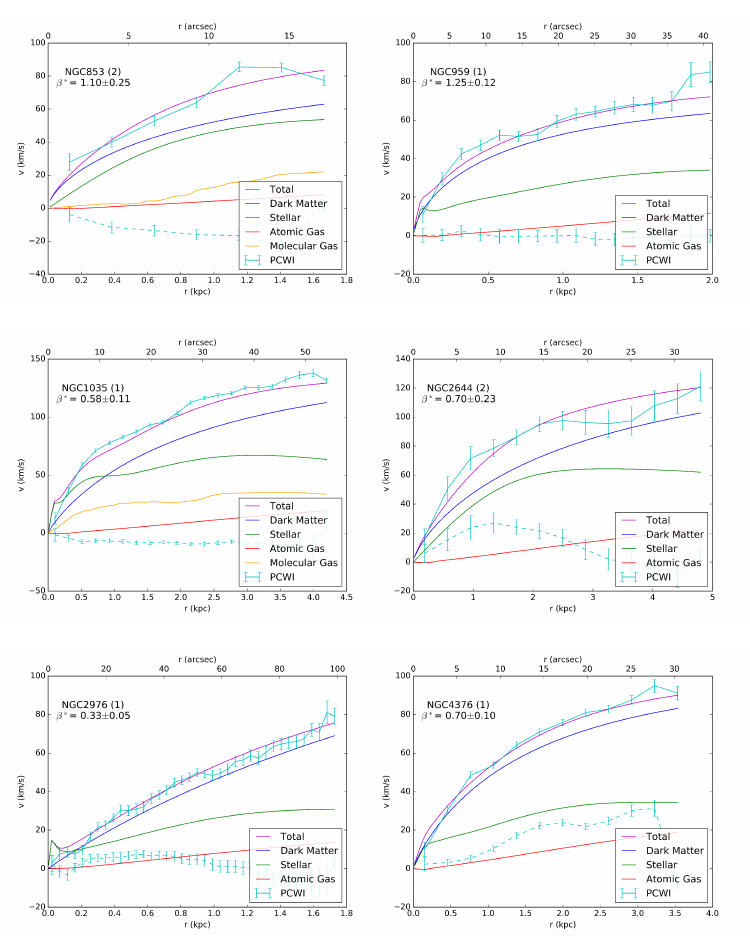}
\end{figure*}
\begin{figure*}
\centering
\includegraphics[width=\textwidth]{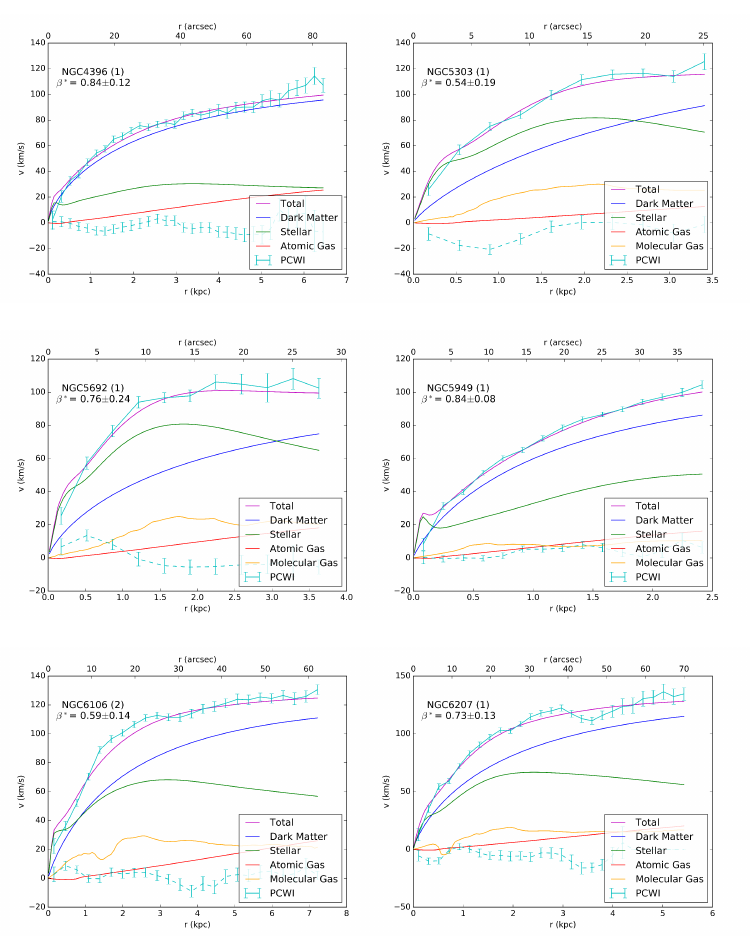}
\end{figure*}

Table \ref{beta_table} gives the values of $\beta^*$ determined for each of our galaxies. We find a mean of $\beta^* = 0.74 \pm 0.07$ and an intrinsic scatter of $0.22^{+0.06}_{-0.05}$. We will discuss the interpretation of the $\beta^*$ distribution and compare it to other observations and theoretical expectations in Sections \ref{results} and \ref{disc}. 

In order to compare this value to that of the NFW profile, we need to calculate the value of $\beta^*$ for a typical galaxy in our mass range if it were to follow NFW exactly. We choose a random sample of 10,000 masses drawn from our galaxies' stellar mass distribution and convert them to halo masses following \cite{Leauthaud}. We then determine the concentrations following the relation and scatter given by Equation 8 of \cite{Dutton}. These are used to construct NFW profiles for each of the 10,000 mock galaxies, which we then use to calculate $\beta^*$. We find that a typical galaxy in our mass range that followed the NFW profile would have a value of $\beta^*_{\rm NFW} = 1.05 \pm 0.02$.


\section{Robustness Tests} 
Since our analysis is based on a number of assumptions based on uncertain models (e.g., the stellar mass-to-light ratio) or incomplete data (e.g., the gas mass distribution) that could potentially have large effects on our results, in this section we will assess the robustness of our inferred dark matter profiles, particularly the value of the inner slope $\beta^*$. 

\subsection{Mass-to-Light Ratio} \label{ML_robust}
Our estimates of $\Upsilon_*$ in Section \ref{ML} rely on the validity of the underlying stellar population synthesis models. In order to test the sensitivity of our dark matter profiles to the estimates of the stellar distribution, we want to explore the full range of kinematically plausible $\Upsilon_*$. As discussed in Section \ref{minmax}, we will consider a minimum and a maximum disk case, which bracket the range of kinematically permitted values of $\Upsilon_*$. 

\begin{figure*}
\centering
\includegraphics[width=\textwidth]{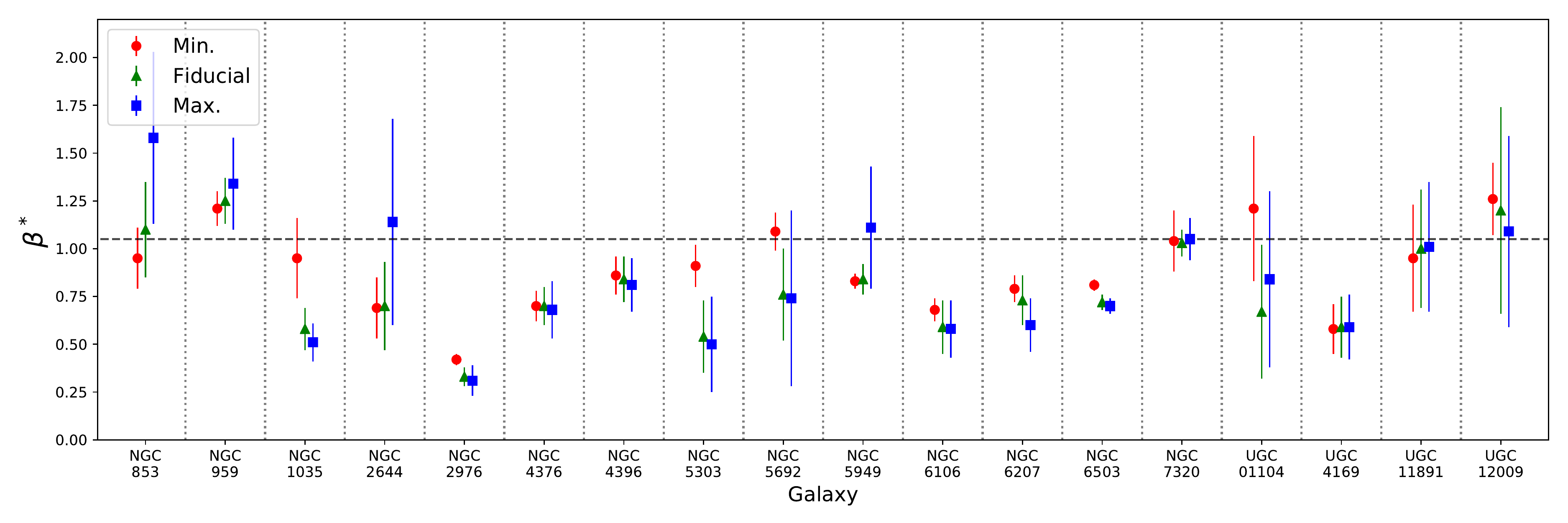} 
\caption{Comparison of values of the inner dark matter slope $\beta^*$ inferred under different assumptions about the stellar mass-to-light ratio $\Upsilon_*$. The fiducial case (green triangles) places a Gaussian prior on $\Upsilon_*$ based on our SPS estimate from Section \ref{ML}, the minimum disk case (red circles) sets $\Upsilon_* = 0$ (dark matter only model), and the maximum disk case (blue squares) fixes $\Upsilon_*$ to the upper bound determined in Section \ref{minmax}. In most cases, there is little to no difference in $\beta^*$ from different the mass-to-light ratios. The horizontal dashed line marks $\beta^*_{\rm NFW} = 1.05$ (see Section \ref{beta*}).}
\label{fig:beta_comp}
\end{figure*}

Figure \ref{fig:beta_comp} compares the value of $\beta^*$ from these two additional cases to our fiducial model for the 18 galaxies with grades 1 or 2. The values of $\beta^*$ for all cases can be found in Table \ref{beta_table}. Typically the three values are close together and fall within one standard deviation of the fiducial value for each galaxy, however there are a few cases where the minimum disk case is larger than the fiducial $\beta^*$ by more than one standard deviation. This is not surprising since this model represents a galaxy with no stellar or gaseous matter whatsoever (dark matter only) and is not expected to be a physically accurate description. 

While in every case, the maximum disk estimate of $\beta^*$ falls within one standard deviation of the fiducial case, it is somewhat surprising to note that the cases that disagree by the largest margin imply steeper profiles than the fiducial case. This is contrary to the usual expectations that the minimum and maximum assumptions bracket the kinematically permitted values of the inner slope. 

We conclude that $\beta^*$ is indeed robust and not strongly influenced by our choice of $\Upsilon_*$, thus any uncertainties in our estimate of $\Upsilon_*$ do not have a significant effect on our final results. We will discuss this insensitivity of $\beta^*$ to $\Upsilon_*$ further in Section \ref{NFWcomp}.

\subsection{Atomic and Molecular Gas}\label{gas_model}
Since the gas mass is sometimes neglected in rotation curve analysis and the relevant data are not always available, we chose to examine the effects of excluding gas from our mass models entirely. We proceed with the same MCMC analysis as before, however we instead model the rotation curves only as $V_{\rm tot}^2 = V_{\rm stars}^2 + V_{\rm DM}^2$. We again keep $\Upsilon_*$ as a free parameter in order to directly compare to our fiducial case. 

We find that the value of $\beta^*$ effectively did not change when we removed the gas from the mass model, the largest difference being 0.09, with 14 of the 18 galaxies having differences of 0.02 or less. 

This is not surprising, as we found the contribution to the rotation curves due to gas to be particularly insignificant in the inner portions, with atomic gas being essentially negligible in the 300-800 pc range in most cases. This leads us to conclude that excluding gas from mass models will not have a tangible effect on the shape of the estimated dark matter distribution, and that galaxies in our sample without molecular gas masses can be considered alongside the others.

\subsection{Kinematic Model}\label{bar-test}
In Section 4.5 of Paper 1, we attempted to fit all of our galaxies with a bisymmetric kinematic model in order to determine if any of our sample was likely to contain a central bar. Only for four of the galaxies (NGC 949, NGC 2976, NGC 3622, NGC 4376) did the bisymmetric fit produce a rotation curve that was significantly different than the axisymmetric fit. For the remaining galaxies there was either no effective difference between the rotation curves produced by the two models, or the bisymmetric model was unable to converge. The median reduced $\chi^2$ value was also lower for the axisymmetric fits, though only somewhat ($\chi^2_{\rm axisym.}/d.o.f. = 0.82,\chi^2_{\rm bisym.}/d.o.f. = 0.97$). 

Closer inspection of the residuals for NGC 3622 showed an unmistakable bar-like pattern, leading us to conclude the galaxy is barred. We thus used the bisymmetric rotation curve for the entirety of our analysis. We could not draw a similar conclusion for the remaining three, so we proceeded with the axisymmetric models for our analysis up until this point. 

To examine what effect the presence of bar might have, we repeat the mass modeling process using the bisymmetric rotation curves for the three candidate galaxies, in addition to the previously discarded axisymmetric fit for NGC 3622. Unfortunately, this new sample of rotation curves was not modeled well by the MCMC procedure (all received a grade of 3; mass models can be seen in Figure \ref{fig:final_plots_3_bar}). The fiducial rotation curves for NGC 949 and NGC 3622 were also not adequately fit by our mass model. Thus we are not able to comment on the effects of the presence of a central bar, however we note that the change to $\beta^*$ was relatively small in all cases. 

The presence of a bar in NGC 2976 was also considered by \cite{N2976} and \cite{Adams} (for a comparison of rotation curves, see Paper 1). \cite{N2976} apply both radial and bisymmetric models to the velocity field of NGC 2976, taken from \cite{Josh03}. They found both fits to be adequate parameterizations of the kinematics, but favor the interpretation that NGC 2976 hosts a bar. \cite{Adams} also model their gas kinematics with both radial and bisymmetric fits, in addition to using stellar kinematics. They find the radial model shows a stronger core than the bisymmetric and stellar models, which are both somewhat steeper. This agreement with stellar kinematic data also leads them to favor the bisymmetric model. While the bisymmetric mass model did not result in a good fit to our data for NGC 2976, we do find the inferred dark matter slope to be slightly steeper than the axisymmetric model ($\beta_{\rm axisym.}^* = 0.33, \beta_{\rm bisym.}^* = 0.42$), consistent with \cite{Adams}. 

Due to our inability to model the bisymmetric rotation curves well, we are unfortunately unable to draw any conclusions about the robustness of $\beta^*$ to the choice of kinematic model for the four galaxies where the kinematic models differed.

\section{Results}\label{results}

\subsection{Distribution of Inner Dark Matter Profile Slopes}\label{dist}
\begin{figure}
\centering
\includegraphics[width=\columnwidth]{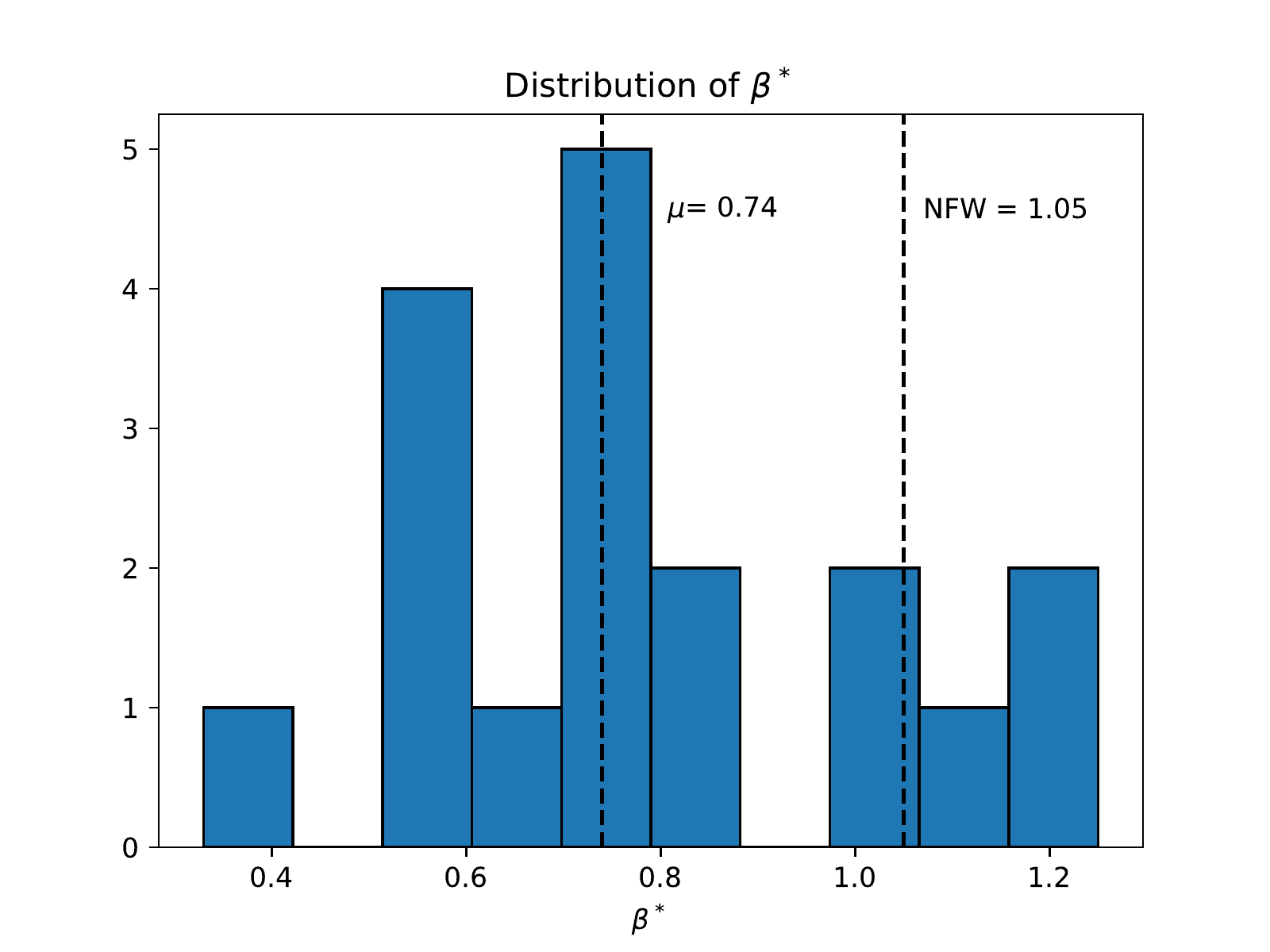} 
\caption{The distribution of inner dark matter density profile slopes (using the fiducial mass-to-light ratio) of the final sample galaxies, which were given mass model grades of 1 or 2. We find one cored galaxy (NGC 2976), several with moderately shallow profiles, and some with NFW-like inner profiles. Our sample has a mean $\beta^*$ of 0.74, while the standard NFW profile corresponds to a $\beta^*$ of 1.05.}
\label{fig:beta_hist}
\end{figure}

Figure \ref{fig:beta_hist} shows the distribution of $\beta^*$ for our sample, while the values are tabulated in Table \ref{beta_table}. Our galaxies cover a range of $\beta^*$, from as shallow as 0.33 to steeper than NFW at 1.25. 

The majority of our galaxies are found to have inner dark matter density profile slopes that are smaller than NFW ($\beta^*$ below 1.05), however many are only moderately shallower, deviating from the expected slope but not enough to be a nearly constant-density core. Around a third are consistent with NFW-like or even steeper profiles. The steeper profiles are likely due to baryonic effects such as adiabatic contraction \citep{Blumenthal, NIHAO9}, but it is notable because any theory put forth to explain the presence of shallow or cored dark matter profiles must also be able to account for steeper profiles in some galaxies. 

We want to quantify the scatter in our distribution, which requires accounting for the different measurement uncertainties for each galaxy. We do this by hypothesizing that the true distribution is Gaussian, then use the ensemble of posteriors on $\beta^*$ from each galaxy to derive constraints on the mean and dispersion of the true distribution. With this method, we find a mean $\langle \beta^* \rangle= 0.74 \pm 0.07$ and intrinsic scatter of $\sigma = 0.22^{+0.06}_{-0.05}$. This $\sigma$ is clearly inconsistent with 0, demonstrating the significant diversity of dark matter profiles in our sample.

To investigate the stability of this mean, we also ran the above calculation on the minimum and maximum disk posteriors. We find $\langle \beta^* \rangle =0.85 \pm 0.06$ and $\langle \beta^* \rangle = 0.70^{+0.08}_{-0.07}$ for the minimum and maximum cases respectively, both of which agree with the fiducial mean within their errors. We also considered only the galaxies that were assigned a grade of 1, finding $\langle \beta^* \rangle = 0.74 \pm 0.09$. The intrinsic scatter shows a similar consistency across the samples. 

Overall, we find that the dark matter profiles in our galaxies are only moderately shallower than NFW, but they have a significant range of values of $\beta^*$, showing that our sample of galaxies contains a diversity of dark matter profiles. This diversity is critical to looking for correlations between the inner slope and other galaxies properties, in addition to determining the cause of these different profiles, as any solution must be able to account for the range we find. 

\subsection{Comparison of Full Rotation Curves to NFW Profiles}\label{NFWcomp}
While our analysis so far has focused on the innermost region (300-800 pc), where the effects of baryonic feedback or new dark matter physics are expected to be strongest, it is also interesting to examine the shape of the rotation curves (and hence the dark matter distribution) as a whole. We would like to explore what, if any, large-scale dark matter properties can be derived from our sample. 

To do this, we first compare the shapes of our baryon-subtracted rotation curves to NFW rotation curves. To determine the shape our rotation curves would take if our galaxies followed the NFW profile, we need to normalize to the maximum rotation velocity for each galaxy, as this value and the mass-concentration relation will determine the parameters in the NFW profile. We cannot always use the maximum velocity taken from our rotation curves as they are often still rising at the outermost measured radius, so we also consider maximum velocities derived from \ion{H}{1} line widths (taken from HyperLeda) and use the larger of the two values. We again use the \cite{Dutton} mass-concentration relation to estimate the value of $M_{200}$ that reproduces the maximum velocity for each galaxy. We then use these NFW profiles to determine the corresponding dark matter contribution to the rotation curve, allowing us to compare directly to the dark matter components inferred from our mass models. 

\begin{figure*}
\centering
\includegraphics[width=\textwidth]{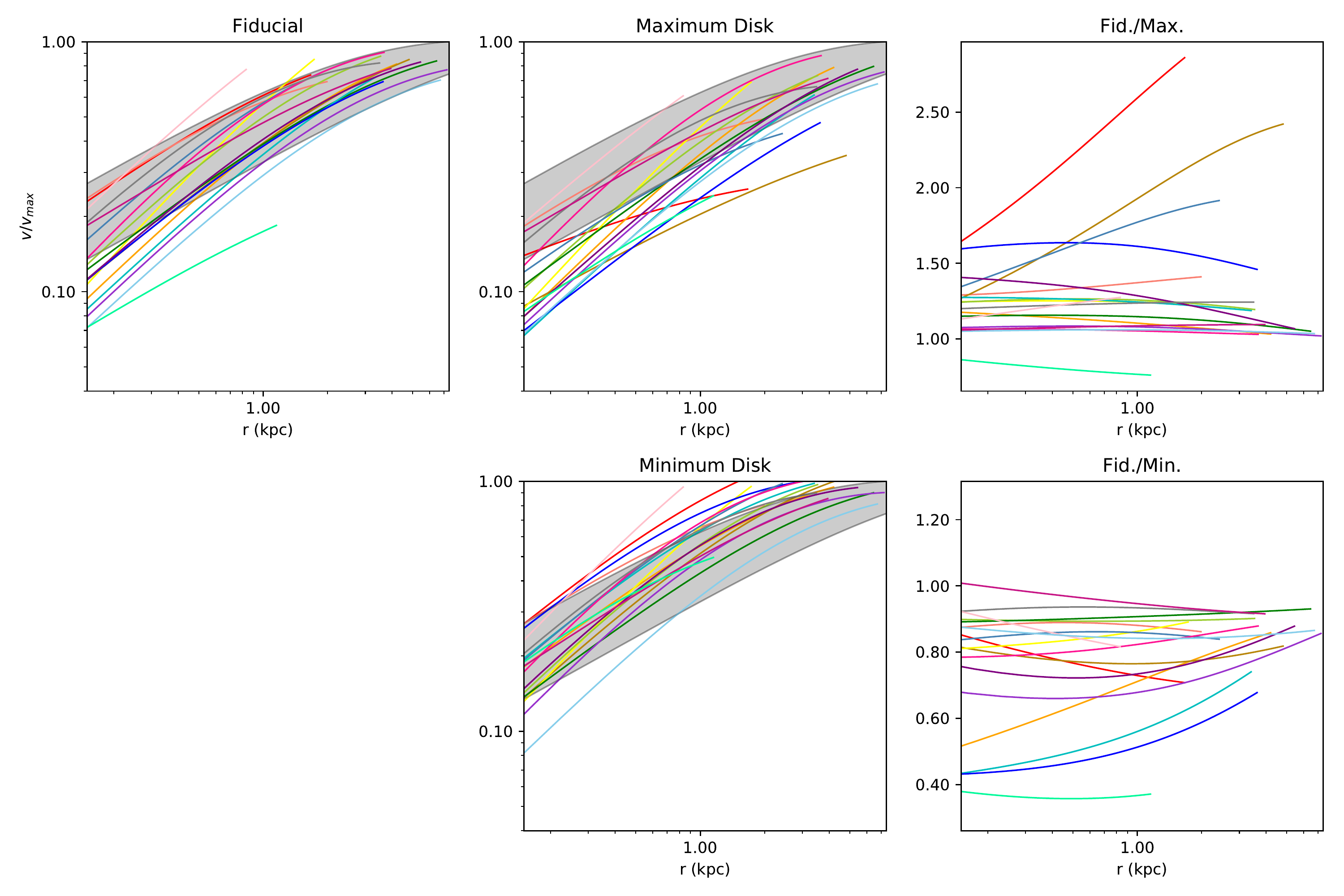} 
\caption{A comparison of the dark matter rotation curves from the fiducial mass models (top left) and the maximum (top middle) and minimum (bottom middle) disk cases, plotted as a fraction of maximum velocity. The gray region is that spanned by NFW profiles on the \cite{Dutton} mass-concentration relation. The right panels give the ratio of the fiducial to maximum (top) and minimum (bottom) disk dark matter curves. For further discussion, see Section \ref{NFWcomp}.}
\label{fig:NFW_comp}
\end{figure*}

The left and center panels of Figure \ref{fig:NFW_comp} show the dark matter contribution to the rotation curves plotted as a fraction of maximum velocity for the fiducial case and the maximum disk assumption, as well as the same for the minimum disk assumption (see Section \ref{minmax}). The curves are overlaid with the region spanned by NFW profiles following the \cite{Dutton} mass-concentration relation. The fiducial case typically has a smaller rotation velocity than NFW in the inner parts ($r \lesssim 1$ kpc), which agrees with our expectation that our galaxies typically have less dark matter in their central regions, then approximately aligns with NFW at large radii, also as expected. The maximum disk case has a similar pattern, and both cases tend to show smaller velocities in the innermost region than NFW, though the fiducial case does so less. In contrast, the minimum disk case aligns well in the inner parts, deviating instead in the outer regions, where some galaxies overestimate the NFW velocities, again in line with expectations as this case assumes all of the galaxies' mass is due to dark matter. 

The right panels of Figure \ref{fig:NFW_comp} give the ratio of the fiducial to maximum and minimum disk dark matter rotation curves. The flat lines in these panels indicate that our sample overall shows similar rotation curve shapes in the fiducial, maximum, and minimum disk cases, despite having different amplitudes. We see that while moving to a maximum or minimum disk model does systematically change the {\it amount} of dark matter in the inner kpc, the {\it slope} in this region does not significantly change in most cases. 

This seems to indicate that insensitivity of $\beta^*$ to the mass-to-light ratio is due to a combination of the relatively small differences between the maximum disk and SPS-predicted values of $\Upsilon_*$ in many cases, and the similarity between the power-law slopes of the stars and dark matter in other cases. 

Note that the only galaxy with a maximum disk ratio less than one is UGC 12009, for which the maximum disk estimate of $\Upsilon_*$ is smaller than the fiducial value. The few galaxies with non-zero slopes in the ratio plot correspond to those with the largest deviations in $\beta^*$ (see Figure \ref{fig:beta_comp}). In these few cases, the rotation curve permits a much larger $\Upsilon_*$ than our SPS-based estimate.  

This indicates that in our sample the dark matter density slope in the inner $\sim 1$ kpc is more robustly measured than the dark matter fraction or concentration. Both depend on the amplitude of the rotation curves and their shape on larger scales, which are covariant with $\Upsilon_*$. Hence in this work, we focus on $\beta^*$ rather than the larger-scale dark matter properties.

\subsection{Correlations}\label{correlations}
The distribution in Figure \ref{fig:beta_hist} shows the variety of density profiles found in our sample of galaxies. This range enables us to look for correlations between the inner slope and other galaxy properties. 

\begin{figure*}
\centering
\includegraphics[width=\textwidth]{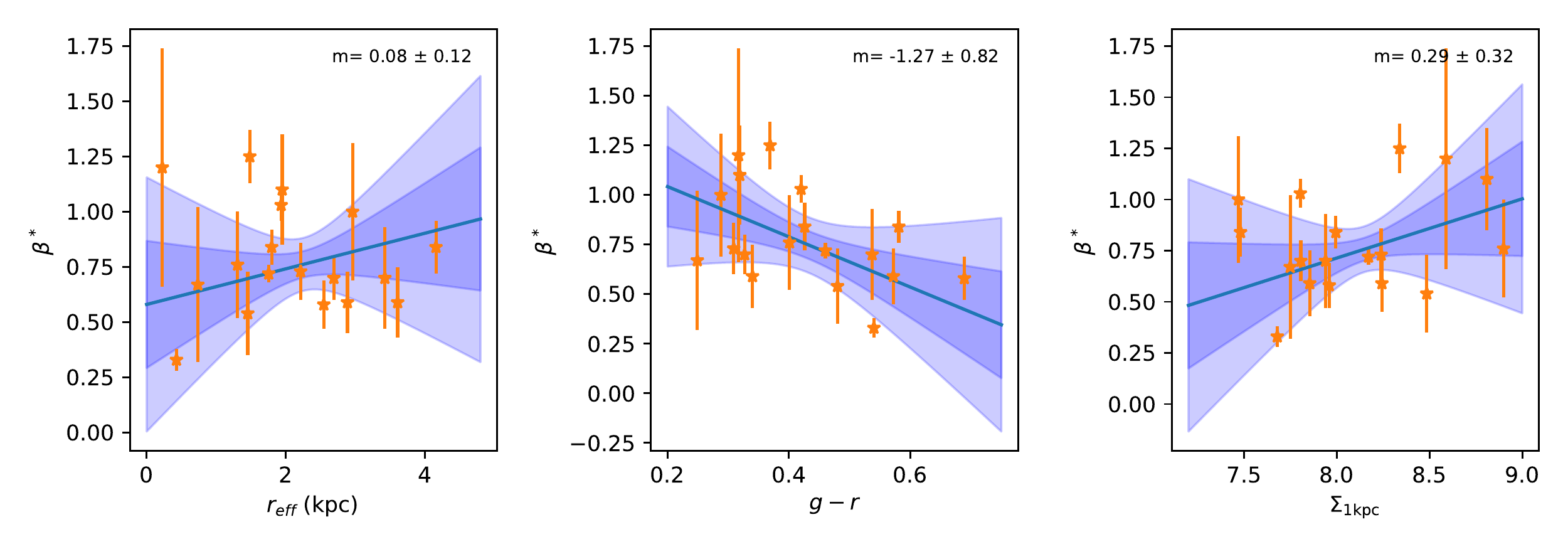}
\caption{Correlations between the inner dark matter slope $\beta^*$ and properties of the stellar populations. The orange stars represent the galaxies in our final sample of galaxies with grades 1 or 2. We use a bootstrap method to estimate the slope parameters and the shaded regions represent 68\% and 95\% confidence levels. The mean slope and 1$\sigma$ uncertainty are given in the upper right corner of each panel. We find no correlation of $\beta^*$ with effective radius or central stellar surface density, however there is tentative evidence of a trend with $g-r$ color, as seen in the middle panel.}
\label{fig:correlation}
\end{figure*}




There are predictions from the Feedback in Realistic Environments (FIRE) simulations \citep{FIRE_OG} that the inner slope of the dark matter profile should correlate with the stellar distribution and specific star formation rate. Strong gas outflows could reduce the gravitational potential, allowing both the dark matter and the stars to migrate outward. This process would be reflected by an increase in effective radius as well as a decrease in specific star formation rate (thus reddening the color; \citealt{FIRE, El-Badry}). \cite{FIRE} also find that in order to maintain a cored profile, a galaxy must have significant late-time star formation to keep the dark matter from re-accreting and forming a cusp. This indicates that core formation may be a continuous process, and the slope of the inner dark matter profile will evolve with time. 

To look for signs of this feedback, we want to examine trends of the dark matter slope with specific star formation rate and effective radius. Figure \ref{fig:correlation} shows the correlations between $\beta^*$ and $r_{\rm eff}$, as well as between $\beta^*$ and $g-r$ color, which we use as a rough proxy for specific star formation rate, as well as the stellar surface density within 1 kpc,  $\Sigma_{1 \rm kpc}$. In the case of $g-r$ color, we find an intriguing suggestion of a correlation with the expected sign based on these physical considerations, however, we are unable to draw firm conclusions because of the low statistical significance of the correlation (1.6$\sigma$). For $r_{\rm eff}$ and $\Sigma_{1 \rm kpc}$, we find no evidence of a correlation. Note that we do not sample low specific star formation rate galaxies, for which tracers other than H$\alpha$ would be required, so this result does not rule out the existence of such a correlation.  

If feedback is responsible for driving dark matter away through fluctuations of the gravitational potential, this mechanism would not be as efficient in more massive galaxies, as their deeper potentials weaken gas outflows. Discussion of a correlation with stellar mass in the context of existing literature as well as a more explicit comparison to the FIRE simulations can be found in Sections \ref{litcomp} and \ref{simulations}.

\section{Discussion}\label{disc}
From our original sample of 26 low-mass galaxies, we were able to adequately model the mass distribution, including dark matter, stars, and gas, in 18 cases. The 18 galaxies in our final sample have stellar masses of $10^{8.1} - 10^{9.7}\msol$ and display a range of inner dark matter slopes. The mean $\beta^*$ is $0.74 \pm 0.07$, while NFW corresponds to $\beta^* = 1.05$, implying that our sample overall has shallower central dark matter profiles than NFW, but only modestly so. Thirteen galaxies in total were determined to have profiles shallower than $\beta^* = 1.05$, while five were consistent with NFW. 

In this section, we will compare these results with those obtained using other kinematic tracers, with observations of dwarf galaxies in the literature, and with simulations of galaxy formation. 

\subsection{Robustness of the Kinematic Tracers}
As discussed previously, \cite{Mai1} used CO velocity fields to produce rotation curves for a sample of dwarf galaxies that overlaps with this work. The second paper in that series, Truong et al. (in prep.), uses the same methodology as this work to produce mass models and analyze the dark matter density profiles, allowing us to make a direct comparison of $\beta^*$ for the seven galaxies that were well fit by both samples. 

\begin{figure*}
\centering
\includegraphics[width=\textwidth]{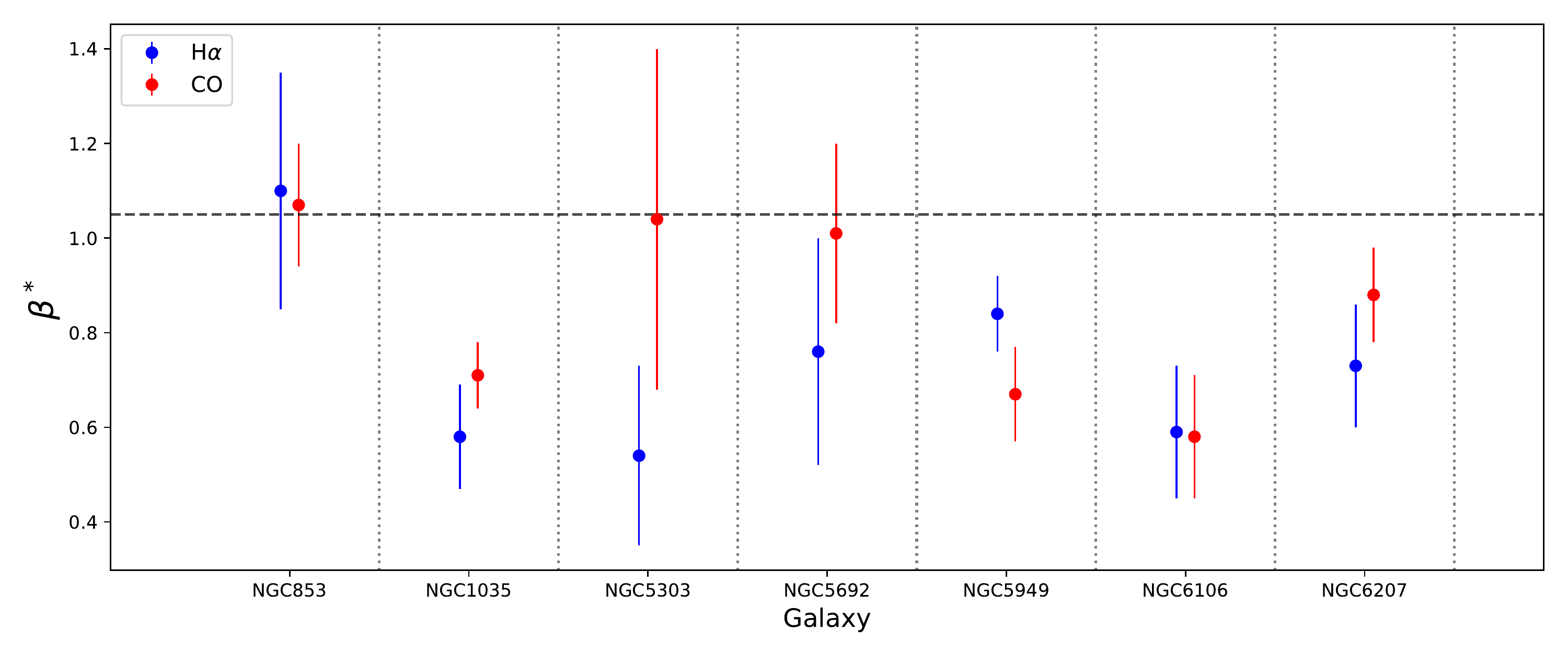}
\caption{The values of the dark matter inner slope $\beta^*$ calculated using H$\alpha$ velocity fields (blue) compared to those using CO velocity fields (red) for the galaxies that overlap with the sample in \cite{Mai1}. There is good agreement between the two kinematic tracers. The horizontal dashed line marks $\beta^*_{\rm NFW}$ = 1.05 (see Section \ref{beta*}).}
\label{fig:CO-Ha}
\end{figure*}

We compare the slopes derived using H$\alpha$ and CO tracers, as seen in Figure \ref{fig:CO-Ha}. Our results are consistent with those from Truong et al. (in prep.) and we do not find any systematic difference between the datasets. Note that although the CO value of $\beta^*$ for NGC 5303 appears to be more in agreement with NFW while the H$\alpha$ shows a core, the CO measurement has a large enough uncertainty that the results are consistent. 

Although one might expect CO to be a more faithful tracer, as it is kinematically colder than the ionized gas, the good agreement we find between CO and H$\alpha$ shows that concerns about ionized gas as a tracer (e.g., outflows from \ion{H}{2} regions, thickness of the ionized gas disk - see \citealt{Levy} for further discussion) do not affect the global kinematics enough to alter estimates of the dark matter profile. 

Since we are tracing gas, our measurements are potentially susceptible to additional outward forces from gas pressure. If this effect is large enough, ignoring it could cause us to underestimate the enclosed mass at a given radius. While approximate corrections for pressure support exist, they depend on several estimations and assumptions. The limitations to the usefulness of these kind of approximations are significant, and are reviewed in detail in \cite{pressureDS}. However, \cite{pressureDS} also indicate pressure support is only likely to be relevant for galaxies with rotation speeds below 75 km s$^{-1}$. Of the galaxies in our final sample, only two (NGC 959 and UGC 12009) fall close to this limit, and in both of these cases we infer a cuspy central density profile. Since correcting for pressure support tends to produce steeper estimates of the density slope, it would not change our overall results for these galaxies. While we cannot draw a stronger conclusion about pressure support without resolved \ion{H}{1} data, we can note that \cite{Adams} used both stellar and gas tracers to infer the dark matter profiles of a similar sample of dwarf galaxies and found general agreement between the two. 

We conclude that the precise choice of tracer (ionized gas, molecular gas, stars) does not significantly affect the derived dark matter profile for late-type galaxies with $M_* \approx 10^9 \msol$.

\subsection{Comparison to Literature Measurements} \label{litcomp}

\begin{figure*}
\centering
\includegraphics[width=\textwidth]{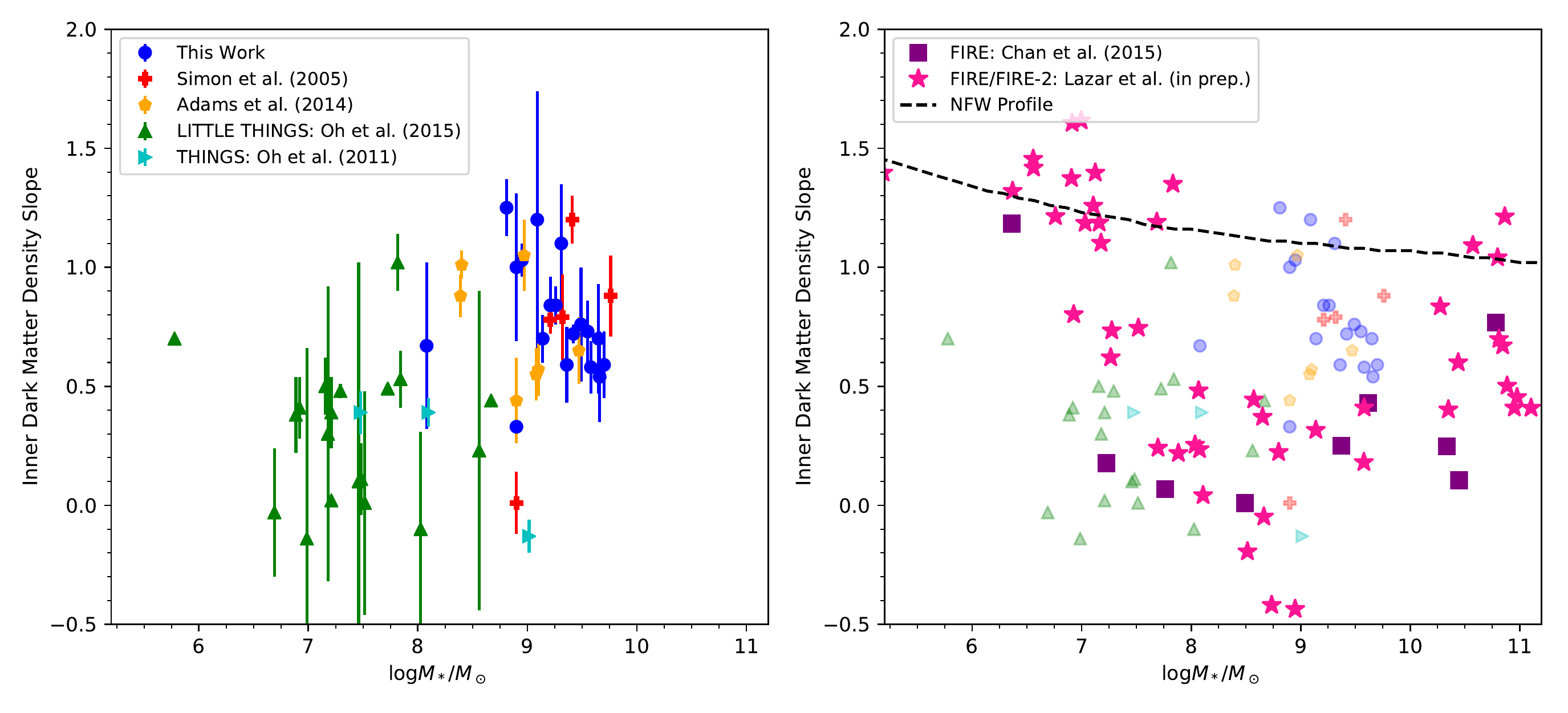}
\caption{The inner dark matter density profile slope versus stellar mass. The left panel shows results from our sample and other similar observational samples in the literature. We have converted the results from \cite{Adams} to $\beta^*$ to agree with our definition of the inner slope. We find evidence for a trend of steepening dark matter profiles in the observational samples over the mass range $M_* \sim 10^7 - 10^{9.5} \msol$. The right panel shows the observations with reduced opacity, overlaid with results from the FIRE and FIRE-2 simulations from \cite{Chan} and Lazar et al. (in prep.). To ensure convergence, we plot only those simulations from the Lazar et al. (in prep.) sample for which the \cite{Power} radius is smaller than 500 pc. The dashed line represents the slope from 300-700 pc (matching the range of \citealt{Chan}) for an NFW profile. The simulations agree reasonably well with observations at lower masses ($M_* \sim 10^8 \msol$), but are too shallow at higher masses ($M_* > 10^{9} \msol$). We note that the precise definition of the inner slope is not the same among the samples, though the simulations from \cite{Chan} and Lazar et al. (in prep.) are measured in a similar range as our $\beta^*$, and the values from \cite{Adams} have been converted to $\beta^*$. See Sections \ref{litcomp} and \ref{simulations} for further discussion.}
\label{fig:comparisons}
\end{figure*}

There is much existing work studying the dark matter density profiles of low-mass galaxies, and we would like to place our results in the context of some of the larger datasets in order to compare and examine potential trends. The left panel of Figure \ref{fig:comparisons} shows the inner dark matter slope versus stellar mass for a union of samples, described briefly here. We note that different groups use different definitions and methodologies for measuring the inner dark matter slope, so the comparison is somewhat imprecise, however we find it to be suitable for examining general trends. 

\cite{Josh05} studied five galaxies using H$\alpha$ velocity fields to derive rotation curves, which they modeled using a power-law dark matter model. They find a mean of $\beta = 0.73$, in excellent agreement with ours. Two galaxies overlap with our sample, NGC 2976 and NGC 5949. For NGC 5949, \cite{Josh05} determines an inner slope of 0.88, which again agrees well with our value of $\beta^* = 0.84$. They found a very shallow inner slope of $0.01 \pm 0.13$ for NGC 2976, which is smaller than our value of $\beta^* = 0.33 \pm 0.05$, however both studies find this galaxy to host the shallowest slope in their respective samples. 

\cite{Adams} studied a sample of seven galaxies and measured both stellar and gas kinematics to derive dark matter density profiles. In order to directly compare to our results, we use their gNFW fits to convert the inner slopes to match our definition of $\beta^*$, finding a mean of $\beta^*_{\rm gas} = 0.74 \pm 0.10$, in excellent agreement with ours. We use these converted values in our discussion and in Figure \ref{fig:comparisons}. The conversion typically increased the slope by around 0.15. We observe three of the same galaxies, allowing us to compare the individual slopes for those cases. NGC 2976 agrees well: \cite{Adams} finds a gas-traced $\beta^*$ of $0.44 \pm 0.18$, while we find $0.33 \pm 0.05$. We find a slope of $0.82 \pm 0.08$ for NGC 5949, which is consistent with the gas-traced slope of $0.65 \pm 0.14$ from \cite{Adams}, as well as with their stellar-traced slope of $0.87 \pm 0.11$. The third galaxy, NGC 959, received a grade of 3 in our analysis and so no comparison of the inner slopes can be made, though we note the rotation curves were in excellent agreement (see Paper 1). 

The THINGS group \citep{Walter} used  \ion{H}{1} observations to study galaxies over a wide range of masses, most of which are beyond the relevant scale for this work. The LITTLE THINGS group followed up THINGS and focused exclusively on 41 dwarf galaxies \citep{Littlethings}. LITTLE THINGS re-observed four of the THINGS dwarf galaxies, so we include just those galaxies from THINGS that were not re-observed \citep{OhTHINGS}. We include the subsample of 26 LITTLE THINGS galaxies studied by \cite{Oh2015}, chosen from the total sample due to the regular rotation patterns in their velocity fields. They find a mean inner dark matter slope of $0.32 \pm 0.24$, indicating much shallower profiles than we find in our sample. 

While \cite{sparc} have constructed mass models for the SPARC data set, which includes dwarfs, we are not able to make a direct comparison as they do not quote logarithmic slopes. \cite{dicintio16} note that there is evidence for cored profiles but better quality rotation curves within 1 kpc would be needed to say more.

Collectively, the observational samples shown in the left panel of Figure \ref{fig:comparisons} span a stellar mass range of $\log M_*/ \msol \approx 7-9.5$ and over this range we find significant evidence for a trend with stellar mass ($p$-value of 0.001). The THINGS and LITTLE THINGS galaxies have shallow slopes, generally clustered around $\beta = 0.5$ and below, while several studies examining higher mass galaxies have found steeper slopes. While there is significant scatter in the inner dark matter profiles at a given mass, the overall trend with mass is still apparent. This correlation provides an important quantitative constraint on dark matter models for future work. 

An important question arises from the heterogeneity amongst the samples, relating to both the precise definition of the inner slope as well as the sample
selection and method of analysis used. It is natural to be concerned that there might be systematic differences that explain some of the trends in Figure \ref{fig:comparisons}.

Regarding the method of measuring the inner slope, our data and the converted data from \cite{Adams} use a definition of the inner slope that is averaged over a particular radial range, namely 300-800 pc. The LITTLE THINGS data, however, fits a power law to a radial range that varies by galaxy, with the inner limit of this range typically falling between 100-400 pc. However, \cite{Oh2015} show that within their sample the inner slope does not correlate with the radial bounds of the fit. This implies that the mean of the LITTLE THINGS data is not sensitive to small differences in this radial range, and so the overall trend with mass cannot be explained by this mild inconsistency. 

In our discussion of the various samples up to this point, we have focused on the difference in stellar mass, but there are other important considerations regarding the observations themselves. In particular, LITTLE THINGS use \ion{H}{1} observations, which require careful treatment of beam smearing, and have velocity fields that are analyzed differently. For instance, they adopt tilted-ring models, in which the geometry of the gas distribution is refit at each radius, rather than fixing the geometric parameters throughout the disk as we do (see Paper 1 for details). There are also physical differences in the velocity fields, which are often more disordered than in higher mass galaxies, and can sometimes lead to the inference of unphysical density profiles (e.g., density increasing with radius in the inner regions). The rotation curves are also analyzed differently. LITTLE THINGS fits a power law to only the innermost few data points (typically 2-3), which tends to minimize the value of the derived slope, whereas we fit a gNFW profile to the full rotation curve then evaluate the slope over a fixed radial range. While it is possible that these differences in methodology could have an influence on the mass trend in Figure \ref{fig:comparisons}, it is beyond the scope of this paper to fully assess that possibility.

\subsection{Comparison to Galaxy Formation Simulations}\label{simulations}
As mentioned previously, the trend with stellar mass seen in the observations in the literature is not entirely unexpected. Qualitatively, we might expect the effects of feedback to be lessened in more massive galaxies, which hints at the possibility of a mass limit above which core creation would be ineffective. It is interesting therefore to make a quantitative comparison with predictions from hydrodynamical simulations. 

The right panel of Figure \ref{fig:comparisons} shows the previously discussed results from observations in reduced opacity, overlaid with results from simulations. We include the results from \cite{Chan}, who use the FIRE simulations to examine the influence of stellar feedback on core creation. They define the inner dark matter slope over the range 300-700 pc, very similar to our range of 300-800 pc used to calculate $\beta^*$. We also include results from Lazar et al. (in prep.), which analyzes simulations run as part of the FIRE and FIRE-2 projects \citep{Fitts, Graus, Wheeler, Chan18-2, El-B17, PH18, G-K, Samuel}. The slopes are measured from 600-800 pc, which is similar to our range but uses a larger lower limit to ensure convergence. The 600-800 pc slopes are 0.2 steeper, on average, than the slopes over 300-800 pc in the best resolved simulations of galaxies around $M_* \sim 10^9$. If we were able to resolve 300-800 pc in all the simulations, better matching the observations, it would likely produce even shallower slopes that only enhance the difference with the observations. The dashed line represents the slope from 300-700 pc (matching the range of \citealt{Chan}) for an NFW profile, calculated as in Section \ref{beta*}. The results of the simulations suggest that the dark matter slope should steepen with stellar mass over the range $M_* \sim 10^7 - 10^{11} \msol$. 


We see a similar trend amongst the overall literature samples. However, while the simulations generally agree with the inner slopes reported for lower mass galaxies  ($M_* \sim 10^8 \msol$), they produce slopes that are, on average, shallower for galaxies in our mass range. There are other properties of the simulated galaxies that differ from our sample. In particular, the simulated galaxies are more diffuse and dispersion supported than typical disk galaxies in this mass range \citep{El-B2018, Chan18}. 

Since strong outflow episodes will produce both shallower dark matter profiles and more diffuse galaxies \citep{El-Badry, dicintio16, Fitts}, these could both be indications that the effects of feedback are too strong in these simulations around $M_* \sim 10^9 \msol$. As simulations improve, it will be interesting to compare the dark matter profiles to galaxies with more similar densities and morphologies. 


The too-shallow slopes found by the FIRE simulations and also the NIHAO simulations \citep{nihao}, which are consistent with FIRE \citep{Bullock}, suggest that the feedback in their models may be too effective around $M_* \sim10^9 \msol$. \cite{Shannon} also challenge the trend between specific star formation rate and effective radius predicted by FIRE for galaxies in the range $M_* \sim 10^9 - 10^{9.5} \msol$, finding the opposite to be true in their observational sample. 

It is also the case that some simulations that include baryonic feedback are unable to produce any shallow dark matter profiles. \cite{eagle}, using the EAGLE simulations, were unable to create cores in dwarf galaxies, as not enough baryons were removed from the central region through feedback. They emphasize the importance of the density threshold for star formation in determining whether or not a core forms.  

Beyond focusing on the inner dark matter density slope, \cite{Oman} emphasized that there is diversity in the shapes of observed rotation curves, even among galaxies with similar $v_{\rm max}$, that is not reflected in the simulations. They emphasize that this dispersion cannot easily be explained through baryon-induced fluctuations to the gravitation potential, calling into question the effectiveness of current simulations. 

The trends in dark matter slopes presented in this paper can be an important constraint (in addition to other observables) on feedback models, which must be able to account for a diversity in galaxy dark matter distributions, morphologies, and star formation rates in this mass range. 

As mentioned in the introduction, SIDM could also explain the shallower profiles observed in dwarf galaxies. Simulations find that shallow dark matter profiles are a natural consequence of SIDM \citep{Elbert, Rocha, Vogel}. \cite{Kamada} were able to use SIDM models to fit a variety of rotation curves with similar maximum velocities but very different shapes in the inner regions. Since our sample of galaxies span a fairly narrow range of stellar masses but show a wide range of inner slopes, it will provide a powerful test of whether SIDM models with a similar cross-section can naturally account for the diversity. We plan to examine this in our future work.

\section{Summary}
The existence of small-scale discrepancies between the dark matter distribution observed in low-mass disk galaxies and that predicted by the $\Lambda$CDM model (in the absence of baryonic effects) has been well established. In this series of papers, we have worked to examine this discrepancy by obtaining high-resolution 2D H$\alpha$ velocity fields from a sample of 26 low-mass galaxies ($M_* = 10^{8.1} - 10^{9.7} \msol$; \citealt{Paper1}) and robustly measuring their inner dark matter density slopes. 

In this paper, we constructed models of our rotation curves as the sum of the stellar, gaseous, and dark matter components. The main conclusions from these mass models are as follows: 

\begin{enumerate}
\item Of the 26 galaxies in our sample, we found 18 were well-suited to measurements of the inner slope of the dark matter density profile. These 18 galaxies had low non-circular motions and our mass models fit the PCWI data and rotation curve shape well. 

\item We found that the quantity that is most robustly constrained by our data is the logarithmic slope $\beta^*$ of the dark matter density profile averaged over the radial range from 300 - 800 pc (Equation \ref{betastar}). This range is motivated by the resolution of our data and that obtained by current simulations. We show that $\beta^*$ is much better constrained by our data than the asymptotic inner slope $\beta$ from the gNFW profile. 

\item The distribution of $\beta^*$ for the fiducial mass models shows a diversity of dark matter profiles in our sample. We find a mean value of $\beta^* = 0.74 \pm 0.07$ and an intrinsic scatter of 0$.22^{+0.06}_{-0.05}$. Around a third of our sample is consistent with NFW-like ($\beta^* = 1.05$) or steeper profiles, while the rest are generally only modestly shallower, with the exception of NGC 2976, which has the flattest density distribution. 

\item Our measurements of the inner dark matter slope were found to be consistent with those derived with CO velocity fields (Truong et al., in prep.), showing that choice of kinematic tracer does not have a significant effect on the derived dark matter profile. The value of $\beta^*$ is also robust to the value of the stellar mass-to-light ratio $\Upsilon_*$, showing that our conclusions are not sensitive to the underlying SPS models. 
 
\item We searched for correlations anticipated based on feedback models (color, effective radius, central stellar surface density). We find no significant trends of $\beta^*$ with $\Sigma_{1 \rm kpc}$ or $r_{\rm eff}$, however we find a suggestion of the expected correlation between $\beta^*$ and color, though the statistical significance is low. 

\item We compared our results to other surveys of dwarf galaxies, finding our sample to agree well with the inner slopes from studies of similar mass galaxies (e.g., \citealt{Josh05}, \citealt{Adams}), and to be generally steeper than LITTLE THINGS, which covers a lower stellar mass range. 

\item Taken together, this indicates an interesting trend of steepening dark matter profiles over the mass range $\log M_*/\msol = 7 - 9$. We discuss the extent to which this trend may be affected by differences in sample selection and analysis techniques.

\item The FIRE simulations, which include feedback models, are in agreement with observations of dwarf galaxies at lower mass scales, however they find inner dark matter slopes that are too shallow in our mass range ($M_* \sim 10^9 \msol$). Supernova feedback thus does not yet provide a quantitative explanation of the dark matter distribution over the full range of dwarf galaxies. 


\end{enumerate}

\section*{Acknowledgements}
We would like to thank Andrew Lazar and James Bullock for providing data used in Figure \ref{fig:comparisons} in advance of publication, T. K. Chan for providing  data used in Figure \ref{fig:comparisons} and for helpful discussions, and Andrew Pontzen and J. A. Sellwood for their insightful comments. 

We acknowledge the usage of the HyperLeda database (http://leda.univ-lyon1.fr). This research has made use of NASA's Astrophysics Data System.

\appendix

\section{Mass Models for Grade 3 Galaxies}
\begin{figure*}
\centering
\includegraphics[width=\textwidth]{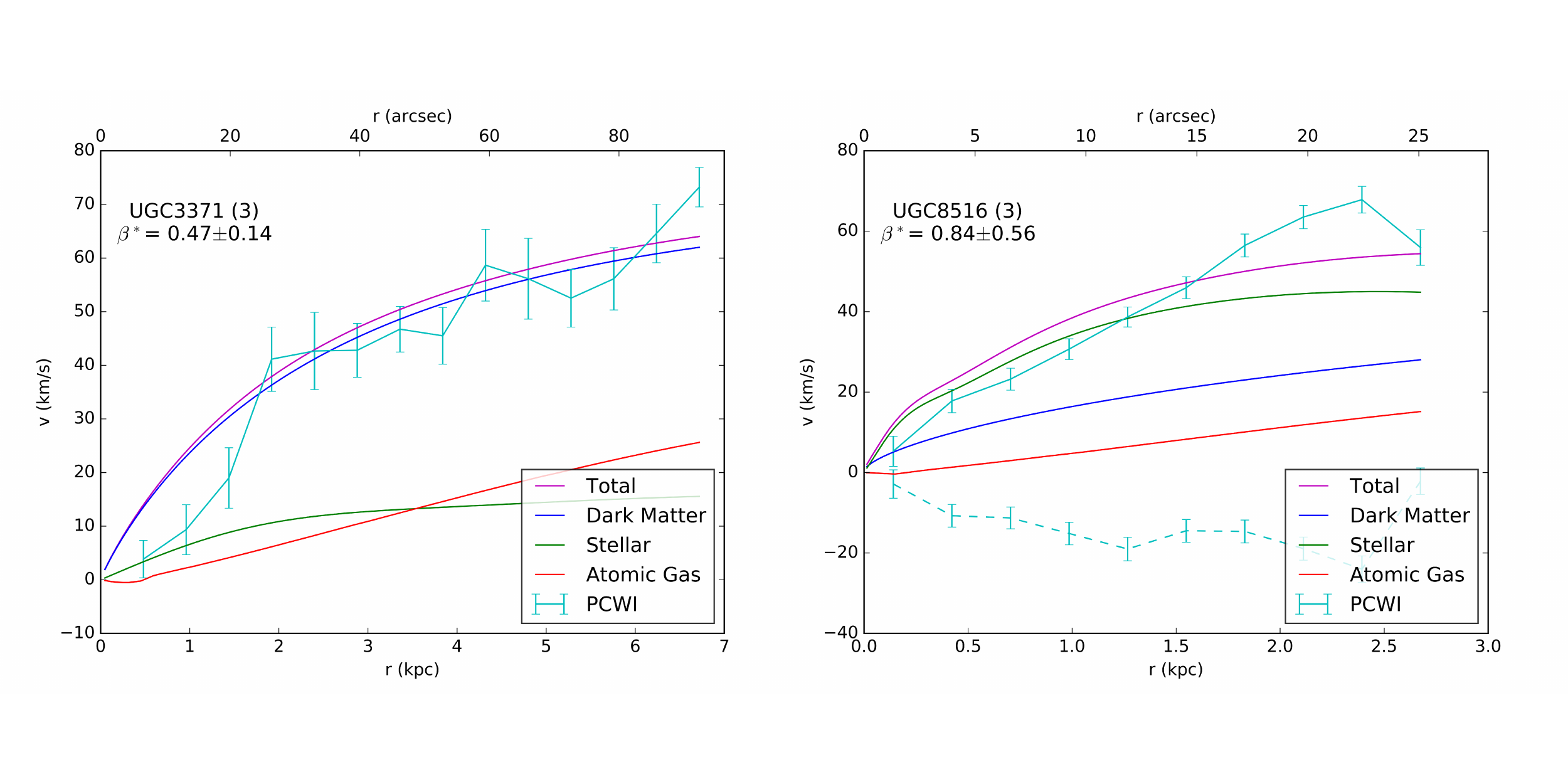}
\caption{Mass models for the galaxies that were graded 3 and not used in further analysis. Each panel contains the galaxy's rotation curve plotted in cyan, derived using data from PCWI. The radial motions are plotted with a dashed line; they are not used in the mass model but are considered in the grading process. Overlaid are the various components of the mass model (total, dark matter, stars, gas). The galaxy name, value of the inner dark matter slope $\beta^*$ and grade are listed in the upper left of each plot. \label{fig:final_plots_3}}
\end{figure*}

\begin{figure*}
\centering
\includegraphics[width=\textwidth]{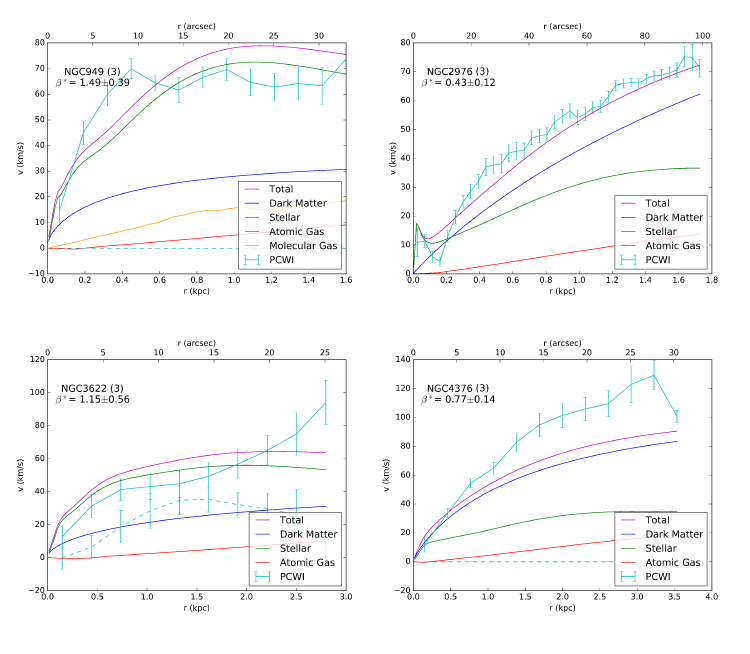}
\caption{Mass models for the galaxies that were well fit with a bisymmetric model (see Section \ref{bar-test}), but for which the resulting mass model received a grade of 3. Note that NGC 3622 was determined to contain a bar (see Paper 1), and so the fiducial case for this galaxy used the bisymmetric model. Here we show the discarded mass model produced from the axisymmetric model. For the remaining three galaxies, we plot the bisymmetric model fits. Each panel contains the galaxy's rotation curve plotted in cyan, derived using data from PCWI. The radial motions are plotted with a dashed line; they are not used in the mass model but are considered in the grading process. Overlaid are the various components of the mass model (total, dark matter, stars, gas). The galaxy name, value of the inner dark matter slope $\beta^*$ and grade are listed in the upper left of each plot. \label{fig:final_plots_3_bar}}
\end{figure*}

Here we present the mass models that were given a grade of 3 and therefore not used for further analysis. Figure \ref{fig:final_plots_3} contains the eight galaxies whose fiducial mass models were given a grade of 3, while Figure \ref{fig:final_plots_3_bar} contains the four mass models from Section \ref{bar-test}, all of which received a grade of 3. These grades were received due to the total model being a poor fit to both the PCWI data and the overall shape of the rotation curve, or due to the presence of significant non-circular motions (see Section \ref{grade} for more details). We note that in same cases (e.g. NGC 949), the amplitude of the rotation curve and model are mismatched. The inclination of the galaxy sets the overall amplitude when constructing the rotation curves, and we use the full covariance matrix from this derivation in our mass models, which can lead to the aforementioned misalignment. 


\begin{figure*}
\centering
\includegraphics[width=\textwidth]{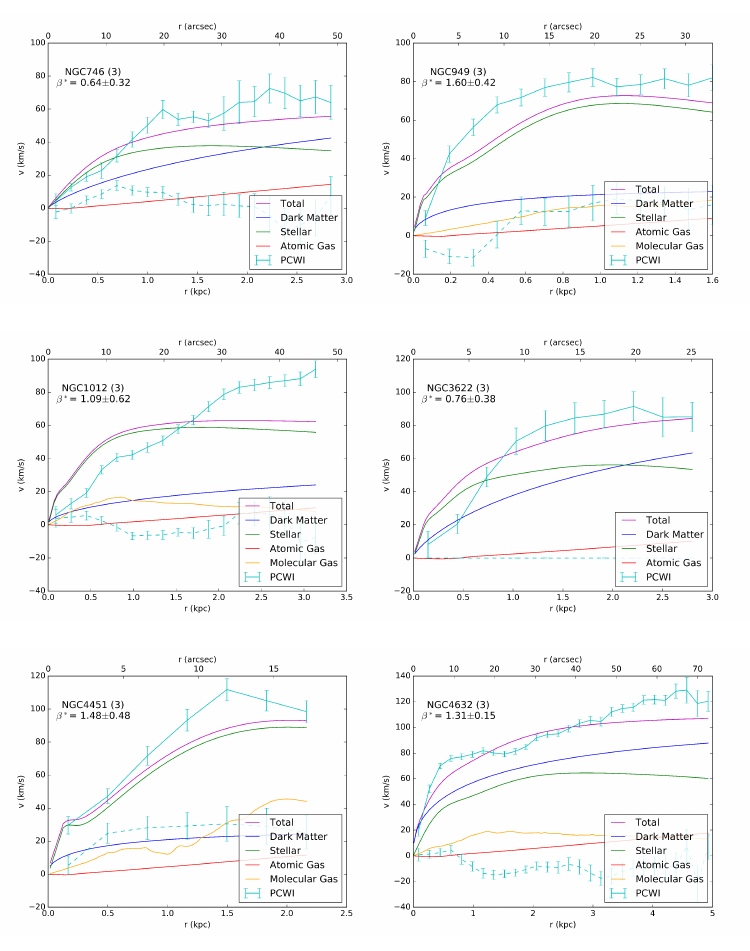}
\end{figure*}

\section{Fiducial MCMC Parameters}
Here we tabulate the remaining MCMC parameters for our fiducial analysis ($\Upsilon_*$ predicted from SPS models). For values of $\beta^*$ and $\Upsilon_*$, see Tables \ref{beta_table} and \ref{ML_table} respectively. We provide this table for completeness, however we caution that because we do not always reach the flat part of the rotation curve, not all of the parameters are well constrained (hence our focus on $\beta^*$). 

\begin{deluxetable}{lcccccc} 
\tablecolumns{7}
\tablewidth{0pt}
\tablecaption{Fiducial MCMC Parameters}
\tablehead{\colhead{Galaxy} & \colhead{$\beta \pm \sigma$} & \colhead{$\log M_{200} \pm \sigma$} & \colhead{$c_{-2}  \pm \sigma$} & \colhead{$\Upsilon_*  \pm \sigma$} &\colhead{$\log j  \pm \sigma$} & \colhead{Grade} }
\startdata
NGC 746 & $\lesssim$ 1.23 & -- & 9.0 $\pm$ 5.2 & 1.05 $\pm$ 0.14 & 0.24 $\pm$ 0.18 & 3  \\ 
NGC 853 & $\lesssim$ 1.29 & 10.8 $\pm$ 0.5 & 23.1 $\pm$ 8.1 & 0.73 $\pm$ 0.19 & 0.94 $\pm$ 0.26 & 2  \\ 
NGC 949 & -- & $\lesssim$ 10.5 & -- & 1.09 $\pm$ 0.13 & 0.73 $\pm$ 0.19 & 3  \\ 
NGC 959 & 0.68 $\pm$ 0.34 & 10.5 $\pm$ 0.4 & $\gtrsim$ 10.8& 0.21 $\pm$ 0.04 & 0.28 $\pm$ 0.19 & 1  \\ 
NGC 1012 & $\lesssim$ 1.76 & -- & $\lesssim$ 29.8 & 0.82 $\pm$ 0.10 & 1.26 $\pm$ 0.18 & 3  \\ 
NGC 1035 & $\lesssim$ 0.61 & 11.5 $\pm$ 0.2 & 19.3 $\pm$ 2.6 & 0.18 $\pm$ 0.03 & 0.50 $\pm$ 0.16 & 1  \\ 
NGC 2644 & $\lesssim$ 0.95 & $\gtrsim$ 10.8 & 15.8 $\pm$ 4.6 & 0.21 $\pm$ 0.04 & 0.02 $\pm$ 0.22 & 2  \\ 
NGC 2976 & $\lesssim$ 0.18 & $\gtrsim$ 11.4 & 16.8 $\pm$ 0.9 & 0.12 $\pm$ 0.02 & -0.07 $\pm$ 0.14 & 1  \\ 
NGC 3622 & $\lesssim$ 1.20 & $\gtrsim$ 9.6 & 15.7 $\pm$ 6.3 & 0.20 $\pm$ 0.04 & 0.44 $\pm$ 0.26 & 3  \\ 
NGC 4376 & $\lesssim$ 0.70 & 11.0 $\pm$ 0.2 & 19.2 $\pm$ 2.8 & 0.21 $\pm$ 0.04 & 0.37 $\pm$ 0.22 & 1  \\ 
NGC 4396 & 0.66 $\pm$ 0.24 & 11.3 $\pm$ 0.2 & 13.6 $\pm$ 3.1 & 0.21 $\pm$ 0.11 & -0.15 $\pm$ 0.16 & 1  \\ 
NGC 4451 & -- & $\lesssim$ 11.2 & -- & 0.22 $\pm$ 0.03 & 0.66 $\pm$ 0.25 & 3  \\ 
NGC 4632 & 1.18 $\pm$ 0.25 & $\gtrsim$ 10.5 & 14.6 $\pm$ 7.4 & 0.21 $\pm$ 0.04 & 0.92 $\pm$ 0.14 & 3  \\ 
NGC 5303 & $\lesssim$ 0.75 & $\gtrsim$ 11.1 & 15.2 $\pm$ 3.0 & 0.18 $\pm$ 0.03 & 0.21 $\pm$ 0.26 & 1  \\ 
NGC 5692 & $\lesssim$ 1.00 & $\gtrsim$ 10.5 & 14.7 $\pm$ 5.3 & 0.77 $\pm$ 0.10 & 0.35 $\pm$ 0.23 & 1  \\ 
NGC 5949 & $\lesssim$ 0.76 & 11.0 $\pm$ 0.2 & 25.0 $\pm$ 3.9 & 0.17 $\pm$ 0.03 & 0.47 $\pm$ 0.21 & 1  \\ 
NGC 6106 & $\lesssim$ 0.71 & 11.3 $\pm$ 0.2 & 17.4 $\pm$ 3.1 & 0.19 $\pm$ 0.03 & 0.57 $\pm$ 0.15 & 2  \\ 
NGC 6207 & $\lesssim$ 0.83 & 11.4 $\pm$ 0.2 & 19.5 $\pm$ 3.8 & 0.21 $\pm$ 0.04 & 0.35 $\pm$ 0.14 & 1  \\ 
NGC 6503 & $\lesssim$ 0.19 & 11.0 $\pm$ 0.4 & 30.0 $\pm$ 1.8 & 0.19 $\pm$ 0.05 & 0.70 $\pm$ 0.12 & 2  \\ 
NGC 7320 & $\lesssim$ 0.76 & 10.7 $\pm$ 0.1 & 30.7 $\pm$ 3.5 & 0.21 $\pm$ 0.04 & 0.18 $\pm$ 0.17 & 1  \\ 
UGC 1104 & $\lesssim$ 0.83 & $\gtrsim$ 9.4 & 18.2 $\pm$ 4.9 & 0.20 $\pm$ 0.04 & 0.41 $\pm$ 0.33 & 2  \\ 
UGC 3371 & $\lesssim$ 0.59 & 10.7 $\pm$ 0.2 & 10.0 $\pm$ 0.0 & 1.06 $\pm$ 0.17 & 0.69 $\pm$ 0.22 & 3  \\ 
UGC 4169 & $\lesssim$ 0.78 & 11.1 $\pm$ 0.2 & 13.6 $\pm$ 2.9 & 0.21 $\pm$ 0.04 & 0.36 $\pm$ 0.17 & 1  \\ 
UGC 8516 & $\lesssim$ 1.59 & -- & 9.3 $\pm$ 7.1 & 0.20 $\pm$ 0.04 & 1.10 $\pm$ 0.24 & 3  \\ 
UGC 11891 & $\lesssim$ 1.44 & 11.1 $\pm$ 0.4 & 15.6 $\pm$ 6.6 & 0.58 $\pm$ 0.16 & 0.07 $\pm$ 0.20 & 2  \\ 
UGC 12009 & $\lesssim$ 1.66 & -- & $\lesssim$ 33.0 & 0.54 $\pm$ 0.10 & 0.19 $\pm$ 0.34 & 2 
\enddata
\tablecomments{Parameter values are determined by the fiducial mass model, which has a Gaussian prior on $\Upsilon_*$ (see Section \ref{MCMC} for details). In several cases, we can provide only an upper or lower limit (95\%), as the posterior distribution peaked at or near the limits of the prior. Similarly, in some cases the posterior distribution was close to flat, and no limits on the parameter could be determined, indicated with a dash.} 
\end{deluxetable}

\bibliographystyle{apj}
\bibliography{bib2}

\end{document}